\documentclass[11pt,a4paper]{article}
\pdfoutput=1

\usepackage{jcappub}
\makeatletter
\def\@fpheader{}
\makeatother
\usepackage{mathrsfs}
\usepackage{amsfonts}
\usepackage{amsthm}
\usepackage{braket}
\usepackage{bm}
\usepackage{caption}
\usepackage{subcaption}
\usepackage{float}
\usepackage{enumerate}
\usepackage{multirow}
\usepackage{array}
\usepackage[figuresright]{rotating}
\usepackage{makecell}
\usepackage{diagbox}
\usepackage{placeins}
\usepackage{booktabs}
\numberwithin{equation}{section}
\numberwithin{table}{section}
\numberwithin{figure}{section}

\title{Background dynamics and cosmological tests of interacting, viscous $H^2+H^{-2}$ dark energy}

\author[a,b,c]{Guo Chen,}
\author[b,c]{Wei Fang}
\author[c,1]{and Chao-Jun Feng\note{Corresponding author.}}

\affiliation[a]{Institute of Fundamental Physics and Quantum Technology, \& School of Physical Science and Technology,\\Ningbo University, Ningbo, Zhejiang 315211, China}
\affiliation[b]{The Shanghai Key Lab for Astrophysics,\\100 Guilin Rd, Shanghai 200234, P. R. China}
\affiliation[c]{Department of Physics, Shanghai Normal University,\\100 Guilin Rd, Shanghai 200234, P. R. China}

\emailAdd{petterchenguo@qq.com}
\emailAdd{wfang@shnu.edu.cn}
\emailAdd{fengcj@shnu.edu.cn}

\abstract{
We study dark-sector energy exchange and bulk viscosity in the HHDE model, whose density combines terms proportional to the square and inverse square of the Hubble parameter.
Five viscosity cases and seven interactions give 35 fluid realizations governed by a common autonomous system.
When the intrinsic dark-energy pressure is reconstructed from the continuity equation, dark-energy viscosity leaves the expansion history unchanged. Interaction and dark-matter viscosity enter through their net contribution, producing further exact background degeneracies.
Each realization admits stable de~Sitter attractors in physical parameter domains, either dominated by dark energy or containing residual dark matter sustained by exchange and viscous stress.
Representative posterior trajectories connect the inferred present states to these endpoints.
Using supernovae, cosmic chronometers, DESI baryon acoustic oscillations and a compressed Planck distance prior, the density-product interaction yields a Hubble constant of $71.15^{+1.53}_{-1.64}$ km s$^{-1}$ Mpc$^{-1}$.
Conditional on the applicability of that prior, the preferred transfer from dark energy to dark matter lowers the minimum chi-squared by 75.16 relative to noninteracting HHDE and reduces the discrepancy with the local distance-ladder measurement to approximately one standard deviation when both uncertainties are combined.
}


\usepackage{adjustbox}
\begin{document}
\maketitle

%===========================================================
\section{Introduction}\label{sec1}
%===========================================================

Late-time cosmic acceleration~\cite{WEINBERG201387} has an economical description in the spatially flat $\Lambda$CDM model~\cite{RevModPhys.75.559}. The unexplained microscopic origin and measured scale of the cosmological constant~\cite{Weinberg2000} have led to dynamical alternatives, including time-dependent scalar fields~\cite{Peebles1988}. Tests beyond $\Lambda$CDM are also motivated by the discrepancy between the Hubble constant inferred from Planck CMB data~\cite{Planck:2018vyg} and that measured with the SH0ES distance ladder~\cite{WOS:000828873600001}. Joint DESI DR2, CMB, and supernova analyses also motivate the study of time-dependent dark energy~\cite{desicollaboration2025desidr2resultsii}.

Holographic dark energy relates the density to an infrared length scale through the holographic principle~\cite{Wang_2017,tHooft:1993dmi}. Saturating the ultraviolet--infrared energy bound gives the scaling $\rho \sim 1/L^2$~\cite{Cohen1999}. Using the future event horizon as the infrared cutoff gives accelerating expansion~\cite{Li_2004}. R\'enyi, Barrow, and Ricci-type generalizations extend the entropy relation or the cutoff prescription~\cite{Sharma_2022,WOS:000749246700001,WOS:000821903600001,NOJIRI2022136844,WOS:000610450200006,RUDRA2023101307}. Studies have also examined viscosity and interactions in the dark sector~\cite{Yang2022,YangJing2023,wang2020,Cataldo:2005qh,WOS:001239390500004}.

The model studied here originates from Kaniadakis holographic dark energy (KHDE). In the Hubble-horizon limit, the leading nontrivial KHDE correction gives the two-parameter $H^{2}+H^{-2}$ dark energy (HHDE) density, $\rho_{DE}=3(\alpha H^2+\widetilde{\beta}H^{-2})$, where $\alpha$ and $\beta=\widetilde{\beta}/H_0^4$ are dark energy correction coefficients.

In our previous work~\cite{ChenGuo_202502}, we constrained HHDE with Pantheon SNe Ia~\cite{Pan-STARRS1:2017jku} and $H(z)$ data~\cite{Sharov:2018yvz}, obtaining $H_0=72.8\,\mathrm{km\,s^{-1}\,Mpc^{-1}}$ and bringing the value inferred from the background closer to the local distance-ladder determination.

Adding viscosity and interaction to HHDE gives 35 Modified $H^{2}+H^{-2}$ Viscous Interacting Dark Energy (MHH-VIDE) fluid realizations ($5$ viscosity cases $\times$ $7$ interaction types). The two DE-viscosity classes are background-degenerate with the no-viscosity class, reducing the background analysis to 21 labeled representatives in Classes 1, 4, and 5; further exact reparameterizations relate interaction and DM viscosity. The resulting degeneracies determine which parameter combinations expansion-history observations can constrain. The observational analysis focuses on noninteracting HHDE and density-product interaction Model 1.7, with Model 5.6 as a dissipative comparison and $\Lambda$CDM and $w$CDM as reference cosmologies.

The late-time attractor conditions and trajectories from the inferred present states test the dynamical consistency of these models. Independent fits of the sound horizon then test how acoustic calibration affects the inferred $H_0$.

Section~\ref{sec2} presents the theoretical framework, Section~\ref{sec3} derives the autonomous systems, and Section~\ref{sec4} analyzes their fixed points and stability. Section~\ref{sec5} presents the observational constraints and model comparison. The Appendices give the complete system in compact form, representative DM-viscosity spectra, the Mathematica implementation, and complete numerical parameter tables.

%===========================================================

\section{Theoretical framework}\label{sec2}
%===========================================================

Kaniadakis entropy is a one-parameter deformation of Boltzmann--Gibbs entropy~\cite{Kaniadakis_2002,Kaniadakis_2005}. Expanding the corresponding holographic energy density to first non-trivial order and choosing the Hubble radius as the infrared cutoff gives~\cite{ChenGuo_202502}
\begin{equation}
\rho_{DE}=3\left(\alpha H^2+\widetilde{\beta}H^{-2}\right),
\label{1.5}
\end{equation}
Here $H=\dot a/a$, $H_0$ is its present value, $E=H/H_0$, and $\beta=\widetilde{\beta}/H_0^4$; a dot denotes a derivative with respect to cosmic time. We use $8\pi G=1$ and restore distance units in the observational analysis.

For a spatially flat FLRW background, the component fractions are $\Omega_i=\rho_i/(3H^2)$, where $i=b,r,DM,DE$ labels baryons, radiation, dark matter, and dark energy. The Friedmann constraint is
\begin{equation}
3H^2=\rho_b+\rho_r+\rho_{DM}+\rho_{DE},\qquad
1=\Omega_b+\Omega_r+\Omega_{DM}+\Omega_{DE}.
\end{equation}
Bulk viscosity contributes an effective pressure \(-3H\xi\). We consider four viscous prescriptions:
\begin{equation}
\xi_{(DE)}=3H\left(\xi_0+\xi_{DE}\Omega_{DE}\right),\qquad
\xi_{(DM)}=3H\left(\xi_1+\xi_{DM}\Omega_{DM}\right),
\label{2.3}
\end{equation}
with at most one of \(\xi_0,\xi_{DE},\xi_1,\xi_{DM}\) nonzero in any model. The five viscosity classes are no viscosity, constant DE viscosity, density-dependent DE viscosity, constant DM viscosity, and density-dependent DM viscosity. Here `constant' refers to the dimensionless coefficient; the bulk-viscosity function contains the factor $H$.

On the expanding, nonnegative-density branch, a nonnegative active viscosity coefficient produces a nonpositive bulk pressure and, at positive temperature, a nonnegative viscous contribution to entropy production~\cite{Cataldo:2005qh}. The $\xi_0$ and $\xi_1$ prescriptions give bulk pressures proportional to $H^2$; the density-dependent prescriptions give $-3\xi_{DE}\rho_{DE}$ and $-3\xi_{DM}\rho_{DM}$, respectively, and vanish with the corresponding density. A zero coefficient removes the viscous stress; negative values describe antidissipation under this constitutive law.

We take the intrinsic pressures to be $p_b=p_{DM}=0$, $p_r=\rho_r/3$, and $p_{DE}=\omega_{DE}\rho_{DE}$. A prime denotes \(d/d\epsilon\), where \(\epsilon=3\ln a\). We write \(Q=9H^3\theta\), where $Q>0$ denotes energy transfer from dark matter to dark energy and $Q<0$ the reverse. The component continuity equations are
\begin{equation}
\begin{aligned}
\rho_b'+\rho_b&=0,\\
\rho_r'+\frac{4}{3}\rho_r&=0,\\
\rho_{DM}'+\rho_{DM}-3H\xi_{(DM)}&=-3H^2\theta,\\
\rho_{DE}'+(1+\omega_{DE})\rho_{DE}-3H\xi_{(DE)}&=3H^2\theta .
\label{2.9}
\end{aligned}
\end{equation}
The effective dark-energy equation of state is
\begin{equation}
\omega_{DE}^{\rm eff}=\omega_{DE}
-\frac{\xi_{(DE)}}{H\Omega_{DE}}.
\label{2.10}
\end{equation}
In this phenomenological closure, we specify Eq.~\eqref{1.5} and the interaction prescription, and Eq.~\eqref{2.9} determines \(\omega_{DE}\). Specifying an independent intrinsic equation of state in addition to \(\rho_{DE}(H)\), \(Q\), and \(\xi_{(DE)}\) would generally overconstrain the background system.

With $\rho_{tot}=3H^2$ and the deceleration parameter $q=-1-\dot H/H^2$, we consider seven interaction prescriptions. The coefficients $\delta$, $\gamma$, and $\eta$ are dimensionless:
\begin{equation}
\begin{aligned}
\theta_1&=\delta\Omega_{DM}+\gamma\Omega_{DE},&
\theta_2&=\delta\Omega_{DM}'+\gamma\Omega_{DE}',\\
\theta_3&=\delta(\Omega_{DM}+\Omega_{DE})
+\gamma(\Omega_{DM}'+\Omega_{DE}'),&
\theta_4&=\gamma,\\
\theta_5&=\gamma\rho_{tot}'/(3H^2),&
\theta_6&=\gamma q,\qquad
\theta_7=\eta\Omega_{DM}\Omega_{DE}.
\end{aligned}
\end{equation}

At the homogeneous-fluid level, the phenomenological linear form $\theta_1$ weights the instantaneous dark-component abundances. The derivative form $\theta_2$ responds to changes in their fractional abundances and vanishes when both fractions are stationary; $\theta_3$ combines the total dark fraction with its rate of change. The constant normalized rate $\theta_4$ gives $Q=3\gamma H\rho_{tot}$, while $\theta_5$ follows the total-density evolution through $Q=\gamma\dot{\rho}_{tot}$. The product in $\theta_7$ switches off the interaction when either dark component is absent.

For $\theta_1$, coefficients of the same sign fix the direction of nonzero transfer, whereas opposite signs allow competing density-weighted contributions. On the expanding physical branch, $\theta_4$ has the sign of $\gamma$, and $\theta_7$ has the sign of $\eta$ when both dark densities are positive. Derivative factors also enter the signs of $\theta_2$ and $\theta_3$. For $\gamma>0$, $\theta_5$ gives DE-to-DM transfer when $\dot H<0$ and DM-to-DE transfer when $\dot H>0$. For the same positive $\gamma$, $\theta_6$ gives DM-to-DE transfer during deceleration ($q>0$) and DE-to-DM transfer during acceleration ($q<0$). Negative $\gamma$ reverses these directions.

%===========================================================
\section{Autonomous systems}\label{sec3}
%===========================================================

The dimensionless phase-space variables are
\begin{equation}
x=\Omega_{DE}=\frac{\rho_{DE}}{3H^2}
=\alpha+\frac{\beta}{E^4},\qquad
y=\Omega_{DM},\qquad z=\Omega_r,
\label{3.1}
\end{equation}
The baryon fraction is \(b=\Omega_b=1-x-y-z\). Differentiating the first relation gives the HHDE identity
\begin{equation}
x'=-2(x-\alpha)\mathcal{G},\qquad
\mathcal{G}\equiv\frac{2H'}{H}.
\label{2.1}
\end{equation}
For $\beta>0$, a finite expanding solution has $x>\alpha$, and $x=\alpha$ is the high-$H$ limiting boundary. Interaction and viscosity affect $x'$ through their effect on the Hubble evolution in $\mathcal G$.

We define
\begin{equation}
v_{DE}\equiv\frac{\xi_{(DE)}}{H}=3\xi_0+3\xi_{DE}x,
\qquad
v_{DM}\equiv\frac{\xi_{(DM)}}{H}=3\xi_1+3\xi_{DM}y.
\end{equation}
Combining Eq.~\eqref{2.9}, the Friedmann equation, and the differentiated HHDE density gives the unified autonomous system
\begin{equation}
\begin{aligned}
\mathcal{G}&=\frac{-1-z/3-\theta+x+v_{DM}}{1+x-2\alpha},\\
x'&=-2(x-\alpha)\mathcal{G},\\
y'&=-y-\theta+v_{DM}-y\mathcal{G},\\
z'&=-\frac{4}{3}z-z\mathcal{G},\qquad
b'=-b-b\mathcal{G}.
\label{3.2}
\end{aligned}
\end{equation}
Differentiating the Friedmann constraint and substituting the component continuity equations yields the same expression for $\mathcal G$. Equation~\eqref{3.2} preserves $x'+y'+z'+b'=0$.

The intrinsic and viscous-effective equations of state of dark energy are
\begin{equation}
\omega_{DE}=-1+\frac{\theta+v_{DE}-x'-x\mathcal{G}}{x},
\qquad
\omega_{DE}^{\rm eff}=-1+\frac{\theta-x'-x\mathcal{G}}{x}.
\label{2.12}
\end{equation}
The pressure that enters the background Einstein equations is
\begin{equation}
p_{DE}^{\rm eff}=p_{DE}-3H\xi_{(DE)}
=3H^2\left(\theta-x'-x\mathcal{G}-x\right).
\label{2.13}
\end{equation}
For fixed \(\rho_{DE}(H)\) and \(Q\), the reconstructed intrinsic pressure compensates the viscous contribution, leaving the effective DE pressure unchanged.
The total effective equation of state and deceleration parameter are
\begin{equation}
\omega_{\rm tot}^{\rm eff}=-1-\mathcal{G},\qquad
q=-1-\frac{3}{2}\mathcal{G}.
\label{3.3}
\end{equation}
The derivative-dependent prescriptions give coupled algebraic equations for $\theta$ and $\mathcal G$. Appendix~\ref{appen.B} gives their explicit reduction.

For each interaction \(j=1,\ldots,7\), Models \(2.j\) and \(3.j\) have \(v_{DM}=0\) and hence the same vector field in Eq.~\eqref{3.2} as Model \(1.j\). At identical shared parameters and initial conditions, they have the same trajectories \(H(a)\), \(x(a)\), \(y(a)\), \(z(a)\), \(q(a)\), \(\omega_{DE}^{\rm eff}(a)\), and \(\omega_{\rm tot}^{\rm eff}(a)\).

Classes 4 and 5 have nonzero \(v_{DM}\) and generally modify the background flow. For some interactions, we can absorb this term by reparameterization, as discussed in Section~\ref{sec5.identifiability}. These further degeneracies reduce the number of independent background models among the 21 labeled representatives.

\subsection{Representative examples}

\subsubsection{\texorpdfstring{Model 1.1: no viscosity and \(\theta=\theta_1\)}{Model 1.1}}\label{Mod1.1}
For \(\xi_{DM}=0\) and \(\theta_1=\delta y+\gamma x\), we define
\[
\mathcal{G}_{1.1}=\frac{-1-z/3-\delta y-\gamma x+x}{1+x-2\alpha}.
\]
Then we obtain
\begin{equation}
x'=-2(x-\alpha)\mathcal{G}_{1.1},\quad
y'=-y-\delta y-\gamma x-y\mathcal{G}_{1.1},\quad
z'=-\frac43z-z\mathcal{G}_{1.1}.
\label{3.1.1.1}
\end{equation}
\begin{equation}
\omega_{\rm tot}^{\rm eff}=-1-\mathcal{G}_{1.1},\qquad
q=-1-\frac32\mathcal{G}_{1.1}.
\label{3.1.1.2}
\end{equation}

\subsubsection{\texorpdfstring{Model 5.1: density-dependent DM viscosity and \(\theta=\theta_1\)}{Model 5.1}}\label{Mod5.1}
For \(\xi_{(DM)}=3\xi_{DM}Hy\), we obtain
\begin{equation}
\begin{aligned}
\mathcal{G}_{5.1}
&=\frac{-1-z/3-\delta y-\gamma x+x+3\xi_{DM}y}{1+x-2\alpha},\\
x'&=-2(x-\alpha)\mathcal{G}_{5.1},\\
y'&=-y-\delta y-\gamma x+3\xi_{DM}y-y\mathcal{G}_{5.1},\\
z'&=-\frac43z-z\mathcal{G}_{5.1}.
\label{3.5.1.1}
\end{aligned}
\end{equation}
\begin{equation}
\omega_{\rm tot}^{\rm eff}=-1-\mathcal{G}_{5.1},\qquad
q=-1-\frac32\mathcal{G}_{5.1}.
\label{3.5.1.2}
\end{equation}

\section{Stability analysis and model screening}\label{sec4}
%===========================================================

\subsection{Methodology}

For each dynamical system, we classify isolated critical points by the eigenvalues of the Jacobian matrix:
\begin{equation}
J(x,y,z) = \begin{pmatrix}
\frac{\partial x'}{\partial x} & \frac{\partial x'}{\partial y} & \frac{\partial x'}{\partial z}\\[4pt]
\frac{\partial y'}{\partial x} & \frac{\partial y'}{\partial y} & \frac{\partial y'}{\partial z}\\[4pt]
\frac{\partial z'}{\partial x} & \frac{\partial z'}{\partial y} & \frac{\partial z'}{\partial z}
\end{pmatrix}.
\label{4.1}
\end{equation}

A critical point is an attractor if all eigenvalues have negative real parts, a repeller if all have positive real parts, and a saddle if the signs are mixed. We solve the fixed-point equations symbolically and substitute exact rational grid parameters before testing physical admissibility and eigenvalues. We discard solutions at poles of the vector field and solutions that are non-real, non-finite, or non-isolated. An accepted numerical attractor must satisfy \(\max_i\operatorname{Re}\lambda_i<-10^{-8}\). This strict criterion excludes non-hyperbolic points and critical lines from the attractor count. The late-time background criterion requires at least one such attractor in the nonnegative-density simplex
\begin{equation}
0\le x\le 1,\quad 0\le y\le 1,\quad 0\le z\le 1,\quad 0\le x+y+z\le 1.
\label{4.2}
\end{equation}
Causal dissipative dynamics, entropy production, and perturbation stability require additional constitutive information.

On the positive-$\beta$ branch, every regular stationary solution with finite $H>0$ has $x_*>\alpha$. Equation~\eqref{2.1} then requires $\mathcal G_*=0$, so that
\begin{equation}
z_*=b_*=0,\qquad y_*=1-x_*,\qquad q_*=-1.
\label{eq:regular_ds}
\end{equation}
These endpoints are de~Sitter. Write $v_{DM}=\nu+\mu y$ and $s=1-\mu$, with $(\nu,\mu)=(0,0)$ for Classes 1--3, $(3\xi_1,0)$ for Class 4, and $(0,3\xi_{DM})$ for Class 5. Using the coefficients $A_j,B_j$ in Appendix~\ref{appen.B}, we define
\begin{equation}
\begin{aligned}
D_*&=1+x_*-2\alpha+B_j(x_*,y_*),\\
\kappa_*&=\left(1-\partial_y v_{DM}-\partial_x A_j+\partial_y A_j\right)_*.
\end{aligned}
\label{eq:regular_coefficients}
\end{equation}
Using Eq.~\eqref{eq:regular_coefficients}, we obtain the three eigenvalues with respect to $\epsilon=3\ln a$:
\begin{equation}
\{\lambda_i\}=\left\{-1,-\frac43,-\frac{2(x_*-\alpha)\kappa_*}{D_*}\right\}.
\label{eq:regular_spectrum}
\end{equation}
For $0<\alpha<1$, a candidate on the algebraic branch of Eq.~\eqref{B.2} is regular, physical, hyperbolic, and attracting if and only if $0\leq y_*<1-\alpha$, $D_*\ne0$, and $\kappa_*/D_*>0$. Table~\ref{tab:regular_ds} lists the candidate coordinates and their coefficients for all seven interactions. These analytic conditions apply throughout the regular branch.

We determine how the critical points depend on the parameters by evaluating them on the finite grids
\begin{equation}
\begin{aligned}
\alpha&\in\{0.033,0.038,0.048,\ldots,0.138\},\\
\gamma,\delta,\eta&\in\{-0.10,-0.09,\ldots,0.10\},\\
\xi_0,\xi_{DE},\xi_1,\xi_{DM}&\in\{0,0.005,0.015,\ldots,0.095,0.10\}.
\label{4.3}
\end{aligned}
\end{equation}

The scan range of $\alpha$ is motivated by the previous KHDE constraint~\cite{ChenGuo_202502}. The interaction-coefficient ranges are symmetric about zero; the viscosity coefficients are nonnegative. Section~\ref{sec5.data} specifies the observational priors, and Appendix~\ref{appen.D} gives the scanning code.

Since the Class-2 and Class-3 vector fields and Jacobians are independent of \(\xi_0\) and \(\xi_{DE}\), we evaluate 422100 parameter combinations for Classes 1, 4, and 5 and use the Class-1 results for Classes 2 and 3.

\subsection{Results by viscosity class}\label{sec4.2}

\subsubsection{Models 1.1--1.7: no viscosity}\label{sec4.2.1}

Each of the seven non-viscous models has a late-time attractor in a nonempty parameter domain. Representative repellers and saddle points also occur within the physical simplex. Table~\ref{tab4.2.1.1} lists the attractor coordinates and existence conditions.

\begin{table}[!htbp]
\centering
\small
\caption{Attractors of Models 1.1--1.7 and their existence conditions within the sampled domain of Eq.~\eqref{4.3}.}
\label{tab4.2.1.1}
\begin{tabular}{ccc}
\toprule
Model & Point B (Attractor) & Existence conditions \\
\midrule
1.1 ($\theta_1$) & $(\frac{\delta+1}{-\gamma+\delta+1},\frac{\gamma}{\gamma-\delta-1},0)$ & $\gamma\leq0$ \\
1.2 ($\theta_2$) & $(1,0,0)$ & All sampled values \\
1.3 ($\theta_3$) & $(\delta+1,-\delta,0)$ & $\delta\leq0$ \\
1.4 ($\theta_4$) & $(1+\gamma,-\gamma,0)$ & $\gamma\leq0$ \\
1.5 ($\theta_5$) & $(1,0,0)$ & All sampled values \\
1.6 ($\theta_6$) & $(1-\gamma,\gamma,0)$ & $\gamma\geq0$ \\
1.7 ($\theta_7$) & $(1,0,0)$ & All sampled values \\
\bottomrule
\end{tabular}
\end{table}

Table~\ref{tab4.2.1.2} lists representative late-time attractors and their Jacobian eigenvalues for $\alpha=0.088$. The equation-of-state column gives the effective DE quantity $\omega_{DE}^{\rm eff}$; every listed attractor has $q=-1$ and hence $\omega_{\rm tot}^{\rm eff}=-1$.

\begin{table}[!htbp]
\centering
\small
\caption{Representative late-time attractors for Models 1.1--1.7 with $\alpha=0.088$.}
\label{tab4.2.1.2}
\resizebox{\textwidth}{!}{%
\begin{tabular}{ccccc}
\toprule
Model (parameters) & Coordinate & Jacobian eigenvalues & $\omega_{DE}^{\rm eff}$ & $q$ \\
\midrule
1.1 ($\gamma=-0.02,\delta=0.01$) & $(0.981,0.019,0)$ & $(-1.333,-1.019,-1.000)$ & $-1.020$ & $-1$ \\
1.2 ($\gamma=-0.02,\delta=0.01$) & $(1,0,0)$ & $(-1.333,-1.000,-0.971)$ & $-1.000$ & $-1$ \\
1.3 ($\gamma=0.02,\delta=-0.01$) & $(0.990,0.010,0)$ & $(-1.333,-1.000,-0.994)$ & $-1.010$ & $-1$ \\
1.4 ($\gamma=-0.02$) & $(0.980,0.020,0)$ & $(-1.333,-1.000,-0.989)$ & $-1.020$ & $-1$ \\
1.5 ($\gamma=-0.02$) & $(1,0,0)$ & $(-1.333,-1.011,-1.000)$ & $-1.000$ & $-1$ \\
1.6 ($\gamma=0.02$) & $(0.980,0.020,0)$ & $(-1.333,-1.006,-1.000)$ & $-1.020$ & $-1$ \\
1.7 ($\eta=0.02$) & $(1,0,0)$ & $(-1.333,-1.020,-1.000)$ & $-1.000$ & $-1$ \\
\bottomrule
\end{tabular}
}
\end{table}

The listed attractor coordinates do not depend explicitly on $\alpha$, although their stability and asymptotic Hubble scale can depend on it through the vector field and $E_*^4=\beta/(x_*-\alpha)$. Here a star denotes an attractor value and $E_*=H_*/H_0$.

The interaction separates these models into two groups. In Models 1.1, 1.3, 1.4, and 1.6, the coupling shifts the attractor away from $(1,0,0)$, with $\omega_{DE}^{\rm eff}<-1$ at the selected couplings. In Models 1.2, 1.5, and 1.7, the attractor remains at $(1,0,0)$ with $\omega_{DE}^{\rm eff}=-1$. For the representative parameter choices, the phase portraits also contain saddle points associated with the radiation-to-matter transition.

\subsubsection{Complete grid-scan results}\label{sec4.2.scan}

Table~\ref{tab:scanconditions} lists the existence conditions on this grid, with accepted and tested parameter counts in parentheses. For each interaction, ``All'' indicates that every tested combination admits an accepted physical attractor.

\begin{table}[!htbp]
\centering
\small
\caption{Conditions for a stable physical attractor on the sampled grid. The Class-1 column also applies to Classes 2 and 3.}
\label{tab:scanconditions}
\resizebox{\textwidth}{!}{%
\begin{tabular}{cccc}
\toprule
$j$ & Classes 1--3 & Class 4: $v_{DM}=3\xi_1$ & Class 5: $v_{DM}=3\xi_{DM}y$ \\
\midrule
1 & $\gamma\leq0$ $(2772/5292)$ & $\gamma\leq3\xi_1$ $(56448/63504)$ & $\gamma\leq0$ $(33264/63504)$ \\
2 & All $(5292/5292)$ & All $(63504/63504)$ & All $(63504/63504)$ \\
3 & $\delta\leq0$ $(2772/5292)$ & $\delta\leq3\xi_1$ $(56448/63504)$ & $\delta\leq0$ $(33264/63504)$ \\
4 & $\gamma\leq0$ $(132/252)$ & $\gamma\leq3\xi_1$ $(2688/3024)$ & $\gamma\leq0$ $(1584/3024)$ \\
5 & All $(252/252)$ & All $(3024/3024)$ & All $(3024/3024)$ \\
6 & $\gamma\geq0$ $(132/252)$ & $\gamma\geq-3\xi_1$ $(2688/3024)$ & $\gamma\geq0$ $(1584/3024)$ \\
7 & All $(252/252)$ & All $(3024/3024)$ & All $(3024/3024)$ \\
\bottomrule
\end{tabular}}
\end{table}

Each of the 35 labeled realizations has a nonempty viable subset on the sampled grid. Models $c.2$, $c.5$, and $c.7$ are unrestricted within this grid for every viscosity class. For Classes 1--3 and 5, Models $c.1$, $c.3$, $c.4$, and $c.6$ require the sign conditions in Table~\ref{tab:scanconditions}. Constant DM viscosity shifts these boundaries through the effective pressure contribution $3\xi_1$: it allows positive $\gamma$ or $\delta$ up to $3\xi_1$ in Models 4.1, 4.3, and 4.4, and negative $\gamma$ down to $-3\xi_1$ in Model 4.6, including equality at each Class-4 boundary. Within Eq.~\eqref{4.3}, the scan found no additional dependence on $\alpha$ or on the unrestricted coupling. Appendix~\ref{appen.C} lists representative attractors and their eigenvalues.

\subsection{Dynamical implications}\label{sec4.3}

Models 1.x--3.x have the same background stability at identical interaction parameters. DM viscosity generally changes the critical points and their eigenvalues; in degenerate cases, exact redefinitions of the interaction coefficients relate the backgrounds. The accepted fractions in Table~\ref{tab:scanconditions} depend on the grid ranges and spacing.

\section{Observational constraints}\label{sec5}
\subsection{Data and parameter inference}\label{sec5.data}

We use the 1048 Type Ia supernovae of the Pantheon compilation, covering $0.01<z<2.3$~\cite{Pan-STARRS1:2017jku}, and the 31 differential-age cosmic-chronometer (CC) measurements compiled by Sharov and Vasiliev~\cite{Sharov:2018yvz}. These data form D1. D2 adds the 13 correlated DESI DR2 BAO observables~\cite{desicollaboration2025desidr2resultsii}; D3 adds the three Planck 2018 TT,TE,EE+lowE distance-prior quantities of Chen, Huang, and Wang~\cite{Chen:2018dbv,Planck:2018vyg}; D4 includes both. The corresponding data counts are 1079, 1092, 1082, and 1095.

We compare flat $\Lambda$CDM, constant-$w$ $w$CDM, noninteracting HHDE, and Models 1.7 and 5.6. Model 1.7, with density-product transfer $\theta=\eta xy$ and zero viscosity, is the principal interacting example. Model 5.6 provides a complementary dissipative example with $\theta=\gamma q$ and $v_{DM}=3\xi_{DM}y$. Both reduce continuously to HHDE when the coupling and viscosity coefficients vanish.

The independent parameters include the supernova absolute magnitude $M$, $H_0$, $\omega_b=\Omega_{b0}h^2$, $\Omega_{c0}=\Omega_{DM,0}$, and $\omega_r=\Omega_{r0}h^2$, together with the model-specific coefficients. Here $h=H_0/(100\,\mathrm{km\,s^{-1}\,Mpc^{-1}})$. For the HHDE family, normalization at $E(0)=1$ gives
\begin{equation}
\begin{aligned}
\Omega_{b0}&=\omega_b/h^2,\\
\beta&=1-\Omega_{c0}-(\omega_b+\omega_r)/h^2-\alpha .
\end{aligned}\label{5.normalization}
\end{equation}
Spatial flatness determines $\beta$; $\Omega_{b0}$ and $\Omega_{M0}=\Omega_{c0}+\omega_b/h^2$ are derived quantities. We compute their posterior intervals from joint samples, retaining parameter correlations.

We adopt uniform priors over $M\in[-25,-15]$, $\omega_b\in[0.005,0.1]$, $\Omega_{c0}\in[0.1,1]$, and $\omega_r\in[0.9,1.1]\omega_{r,\mathrm{ref}}$, with $\omega_{r,\mathrm{ref}}=2.469\times10^{-5}(1+0.2271\times3.046)$. The $H_0$ ranges are $[20,100]$ for $\Lambda$CDM, $w$CDM, and HHDE, and $[40,100]\,\mathrm{km\,s^{-1}\,Mpc^{-1}}$ for the interacting extensions. Additional ranges are $w\in[-3,0]$, $\alpha\in[0.001,0.5]$, $\eta\in[-0.3,0.3]$, and $\xi_{DM}\in[0,0.1]$. Model 5.6 uses $\gamma\in[0,0.3]$ in D1 and $[-0.3,0.3]$ in D4. The signed D4 domain also requires positive $\beta$ and regular, nonnegative background densities over the past integration interval. Comparing D1 and D4 therefore changes both the data and the allowed coupling range. We specify the observational priors independently of the finite dynamical grid.

Background distances follow from the model expansion history; the sound horizons are calculated using CAMB~\cite{Lewis:1999bs}. The calculation fixes $T_{\rm CMB}=2.7255\,\mathrm{K}$ and $Y_{\rm He}=0.2454$, treats neutrinos as massless, and derives their effective radiation contribution from the sampled $\omega_r$. The recombination history is calculated with the CAMB RECFAST option calibrated to CosmoRec~\cite{Seager1999,ChlubaThomas2011,Lewis2026CAMB}. An equivalent dark fluid with the combined DE and DM density represents their expansion history. The D3 and D4 constraints depend on the applicability of the compressed Planck information to these interacting fluids.

We retain the released Pantheon, DESI, and Planck-prior covariances and the quoted CC uncertainties. We sample the Gaussian likelihood with an affine-invariant ensemble algorithm~\cite{ForemanMackey2013}. The D4 chains for Models 1.7 and 5.6 satisfy the primary two-successive-evaluation criterion. Appendix~\ref{appen.E} reports the supplementary independent-pair checks; daggers identify cases with $\widehat R_{40}\geq1.01$. Posterior summaries are medians and central $68.3\%$ intervals. Bounded multistart optimization over the parameters, including $M$, gives the joint numerical likelihood maxima.

\paragraph{AI-assisted computation.}
DeepSeek assisted with Python debugging, and GPT-6 with analytic calculations, consistency checks, and post-processing and plotting scripts. The symbolic and numerical checks are described in Appendices~\ref{appen.D} and~\ref{appen.E}. The figures were generated by plotting numerical outputs.

\subsection{Background identifiability of dark-sector parameters}\label{sec5.identifiability}

The degeneracy derived in Section~\ref{sec3} implies
\begin{equation}
\mathcal L_{2.j}(\boldsymbol\vartheta,\xi_0)=\mathcal L_{3.j}(\boldsymbol\vartheta,\xi_{DE})=\mathcal L_{1.j}(\boldsymbol\vartheta),
\label{5.degeneracy}
\end{equation}
where $\boldsymbol\vartheta$ contains the shared parameters. With factorized priors, the marginal posterior of either DE-viscosity coefficient equals its prior. Expansion data cannot identify these coefficients, so they do not count as additional identifiable background parameters.

Interaction and DM viscosity enter Eq.~\eqref{3.2} through $\mathcal I=\theta-v_{DM}$. Model 4.4 is background-equivalent to Model 1.4 at $\gamma_{\rm eff}=\gamma-3\xi_1$, and Model 5.1 to Model 1.1 at $\delta_{\rm eff}=\delta-3\xi_{DM}$ with unchanged $\gamma$. Equal marginalized predictions require matching the induced priors on these effective combinations.

\subsection{Posterior constraints and data dependence}\label{sec5.results}\label{sec5.datacombinations}

\begin{figure}[!htbp]\centering
\includegraphics[width=\linewidth]{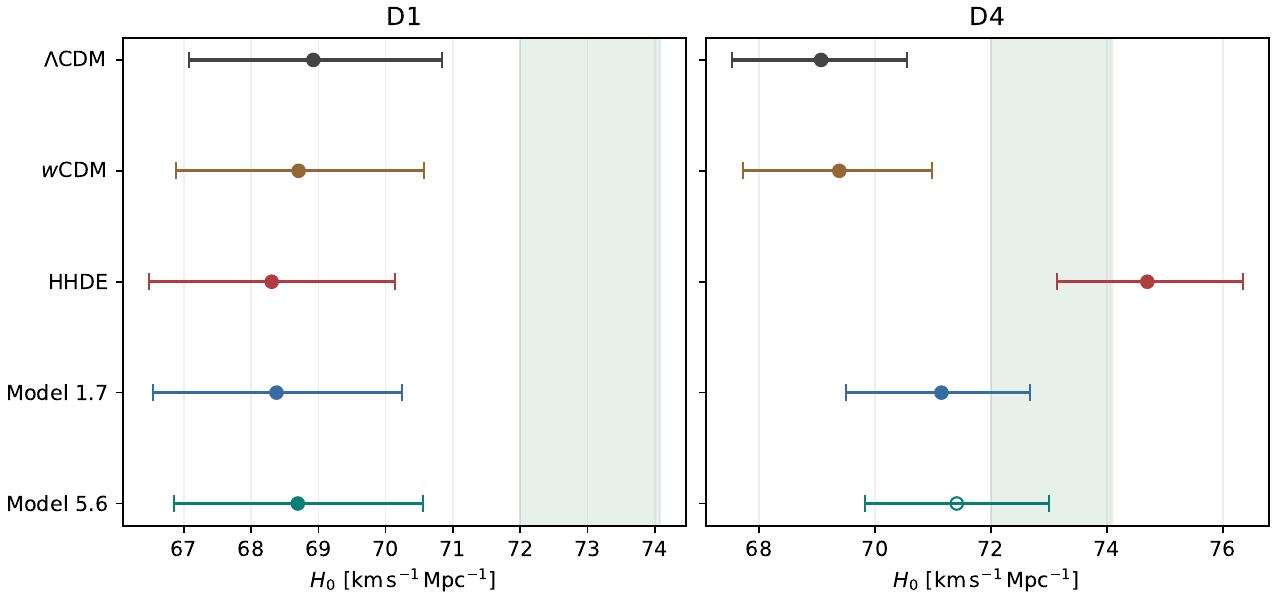}
\caption{$H_0$ summaries for D1 and D4. The open symbol denotes the borderline supplementary pairwise diagnostic for Model 5.6 D4. The shaded band shows the SH0ES value $73.04\pm1.04\,\mathrm{km\,s^{-1}\,Mpc^{-1}}$~\cite{WOS:000828873600001} as an external comparison.}\label{fig:dataset_H0}
\end{figure}

Table~\ref{tab:fit1} summarizes the expansion parameters and total matter density. D1 places all five $H_0$ medians in the range 68.31--68.93 km s$^{-1}$ Mpc$^{-1}$, with uncertainties near 1.8--1.9 km s$^{-1}$ Mpc$^{-1}$. With D4, HHDE gives $H_0=74.70^{+1.65}_{-1.56}\,\mathrm{km\,s^{-1}\,Mpc^{-1}}$, and the reference models remain near 69. Model 1.7 yields $71.15^{+1.53}_{-1.64}\,\mathrm{km\,s^{-1}\,Mpc^{-1}}$; Model 5.6 gives $71.41^{+1.59}_{-1.58}\,\mathrm{km\,s^{-1}\,Mpc^{-1}}$. Figure~\ref{fig:dataset_H0} shows how the added acoustic information changes these estimates.

\begin{table}[!htbp]\centering\small
\caption{Medians and central $68.3\%$ intervals; $H_0$ is in km s$^{-1}$ Mpc$^{-1}$. A dagger denotes a borderline supplementary pairwise diagnostic; see Appendix~\ref{appen.E}.}\label{tab:fit1}
\begin{tabular}{llcc}
\toprule
Model & Data & $H_0$ & $\Omega_{M0}$\\
\midrule
$\Lambda$CDM & D1 & $68.93^{+1.91}_{-1.84}$ & $0.3028^{+0.0204}_{-0.0197}$\\
$w$CDM & D1 & $68.71^{+1.86}_{-1.82}$ & $0.3274^{+0.0418}_{-0.0435}$\\
HHDE & D1 & $68.31^{+1.83}_{-1.83}$ & $0.3226^{+0.0506}_{-0.0671}$\\
Model 1.7 & D1 & $68.38^{+1.86}_{-1.83}$ & $0.3292^{+0.0571}_{-0.0621}$\\
Model 5.6 & D1 & $68.70^{+1.86}_{-1.84}$ & $0.4460^{+0.1025}_{-0.1003}$\\
$\Lambda$CDM & D4 & $69.07^{+1.48}_{-1.54}$ & $0.3011^{+0.0038}_{-0.0038}$\\
$w$CDM & D4 & $69.38^{+1.60}_{-1.65}$ & $0.2985^{+0.0057}_{-0.0056}$\\
HHDE & D4 & $74.70^{+1.65}_{-1.56}$ & $0.2630^{+0.0033}_{-0.0032}$\\
Model 1.7 & D4 & $71.15^{+1.53}_{-1.64}$ & $0.2840^{+0.0046}_{-0.0045}$\\
Model 5.6$^{\dagger}$ & D4 & $71.41^{+1.59}_{-1.58}$ & $0.2842^{+0.0046}_{-0.0045}$\\
\bottomrule
\end{tabular}

\end{table}

Table~\ref{tab:fit2} shows that D1 leaves $\alpha$ broadly distributed. D4 localizes it near 0.020 in HHDE, at $0.2358^{+0.0163}_{-0.0171}$ in Model 1.7, and at $0.2213^{+0.0264}_{-0.0258}$ in Model 5.6.

\begin{table}[!htbp]\centering\small
\caption{HHDE density parameters and dark-sector parameters. We derive $\beta$ from joint samples. The dagger has the same meaning as in Table~\ref{tab:fit1}.}\label{tab:fit2}
\begin{adjustbox}{max width=\linewidth}\begin{tabular}{llccccc}
\toprule
Model & Data & $\alpha$ & $\beta$ & $\gamma$ & $\xi_{DM}$ & $\eta$\\
\midrule
HHDE & D1 & $0.1794^{+0.1719}_{-0.1255}$ & $0.4954^{+0.0777}_{-0.1046}$ & -- & -- & --\\
Model 1.7 & D1 & $0.1907^{+0.1739}_{-0.1330}$ & $0.4730^{+0.0931}_{-0.1150}$ & -- & -- & $-0.1097^{+0.2133}_{-0.1368}$\\
Model 5.6 & D1 & $0.1653^{+0.1690}_{-0.1175}$ & $0.3667^{+0.0886}_{-0.0903}$ & $0.0843^{+0.0988}_{-0.0603}$ & $0.0675^{+0.0225}_{-0.0313}$ & --\\
HHDE & D4 & $0.0204^{+0.0057}_{-0.0057}$ & $0.7165^{+0.0064}_{-0.0065}$ & -- & -- & --\\
Model 1.7 & D4 & $0.2358^{+0.0163}_{-0.0171}$ & $0.4800^{+0.0198}_{-0.0191}$ & -- & -- & $-0.1361^{+0.0075}_{-0.0071}$\\
Model 5.6$^{\dagger}$ & D4 & $0.2213^{+0.0264}_{-0.0258}$ & $0.4943^{+0.0279}_{-0.0282}$ & $0.0144^{+0.0100}_{-0.0090}$ & $0.0143^{+0.0029}_{-0.0025}$ & --\\
\bottomrule
\end{tabular}
\end{adjustbox}
\end{table}

D2 and D3 show how adding BAO or the compressed CMB prior separately changes the HHDE constraints. D2 yields $H_0=72.31^{+1.69}_{-1.70}$ and D3 yields $68.48^{+1.80}_{-1.69}$ km s$^{-1}$ Mpc$^{-1}$, with $\alpha=0.1258^{+0.1183}_{-0.0889}$ and $0.0826^{+0.0091}_{-0.0095}$, respectively. BAO raises the preferred expansion scale, and the compressed CMB information alone chiefly localizes $\alpha$. The estimates are correlated because they share data. D3 passes the primary sampling criterion; its supplementary pairwise diagnostic is marginal, as reported in Appendix~\ref{appen.E}.

\begin{figure}[!htbp]\centering
\includegraphics[width=\linewidth]{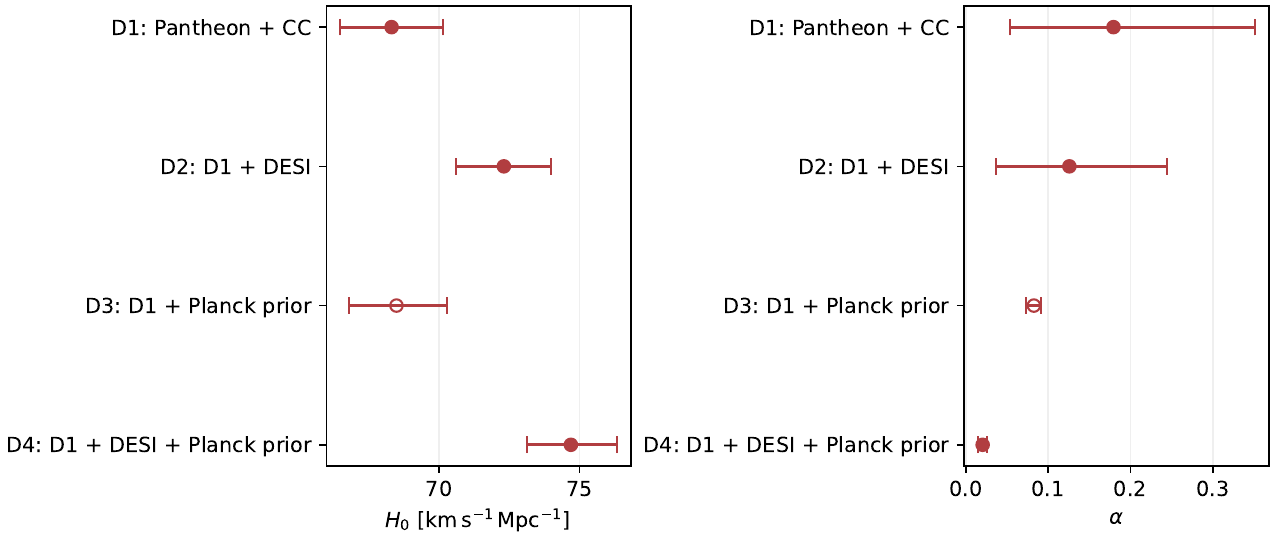}
\caption{HHDE results for the four data combinations. Bars show central $68.3\%$ intervals. The open D3 symbol indicates its borderline supplementary pairwise sampling diagnostic. D2 uses the predicted sound horizon; the test in Section~\ref{sec5.calibration} fits $r_d$ independently.}\label{fig:HHDEprobes}
\end{figure}

\subsection{Dynamical consistency of the inferred backgrounds}\label{sec5.medianstability}

For Model 1.7,
\begin{equation}
\begin{aligned}
(x_*,y_*,z_*)&=(1,0,0),\\
\{\lambda_i\}&=\{-1,-4/3,-(1+\eta)\},\\
E_*^4&=\beta/(1-\alpha).
\end{aligned}\label{eq:17attractor}
\end{equation}
Positive $\beta$, $\alpha<1$, and $\eta>-1$ suffice for local stability. The D1 and D4 medians give interaction eigenvalues of approximately $-0.8903$ and $-0.8639$. These inferred states satisfy the analytic local-stability conditions.

For Model 5.6, set $s=1-3\xi_{DM}$ and $D_*=1+x_*-2\alpha-3\gamma/2$. The de~Sitter critical point has
\begin{equation}
\begin{aligned}
(x_*,y_*,z_*)&=(1-\gamma/s,\gamma/s,0),\\
\{\lambda_i\}&=\{-1,-4/3,-2s(x_* -\alpha)/D_*\}.
\end{aligned}\label{eq:56attractor}
\end{equation}
Conditions $0\leq\xi_{DM}<1/3$, $0\leq\gamma<s$, $\alpha<x_*$, and $D_*>0$ suffice for a stable physical endpoint on the positive-$\beta$ branch, with $E_*^4=\beta/(x_* -\alpha)$. The nonnegative-coupling prior box used for D1 satisfies these local conditions whenever $\beta>0$. Within the signed D4 box, negative $\gamma$ gives $y_*<0$ and violates the physical-endpoint condition. These local conditions require a separate check from past-background regularity. Both reported median parameter vectors satisfy the local endpoint conditions.

\subsection{Posterior trajectories and energy exchange}\label{sec5.supplement}

For each of HHDE, Model 1.7, and Model 5.6 under D1 and D4, we draw 400 joint parameter vectors uniformly without replacement and integrate their trajectories from the inferred present states to the late-time endpoints.

With $u=\ln a$ and $a=1$ today, the HHDE identity and separate conservation of baryons and radiation give
\begin{equation}
\begin{aligned}
\frac{d\ln E^2}{du}&=3\mathcal G,\qquad x=\alpha+\beta/E^4,\\
\Omega_b&=\Omega_{b0}e^{-3u}/E^2,\\
\Omega_r&=\Omega_{r0}e^{-4u}/E^2,\\
y&=1-x-\Omega_b-\Omega_r .
\end{aligned}\label{eq:trajectoryclosure}
\end{equation}
The corresponding prescription in Eq.~\eqref{3.2} defines $\mathcal G$. We integrate from redshift $10^4$ to $a=10^3$ with relative and absolute tolerances of $2\times10^{-10}$ and $2\times10^{-12}$, respectively. We check the component densities at 701 output points with a numerical tolerance of $2\times10^{-7}$. All 2400 selected histories span this interval and undergo a single deceleration-to-acceleration transition. At $a=10^3$, the largest endpoint error $\max\{|x-x_*|,|y-y_*|,|q+1|,|E/E_*-1|\}$ is $5.1\times10^{-7}$.

\begin{figure}[!htbp]\centering
\includegraphics[width=\linewidth]{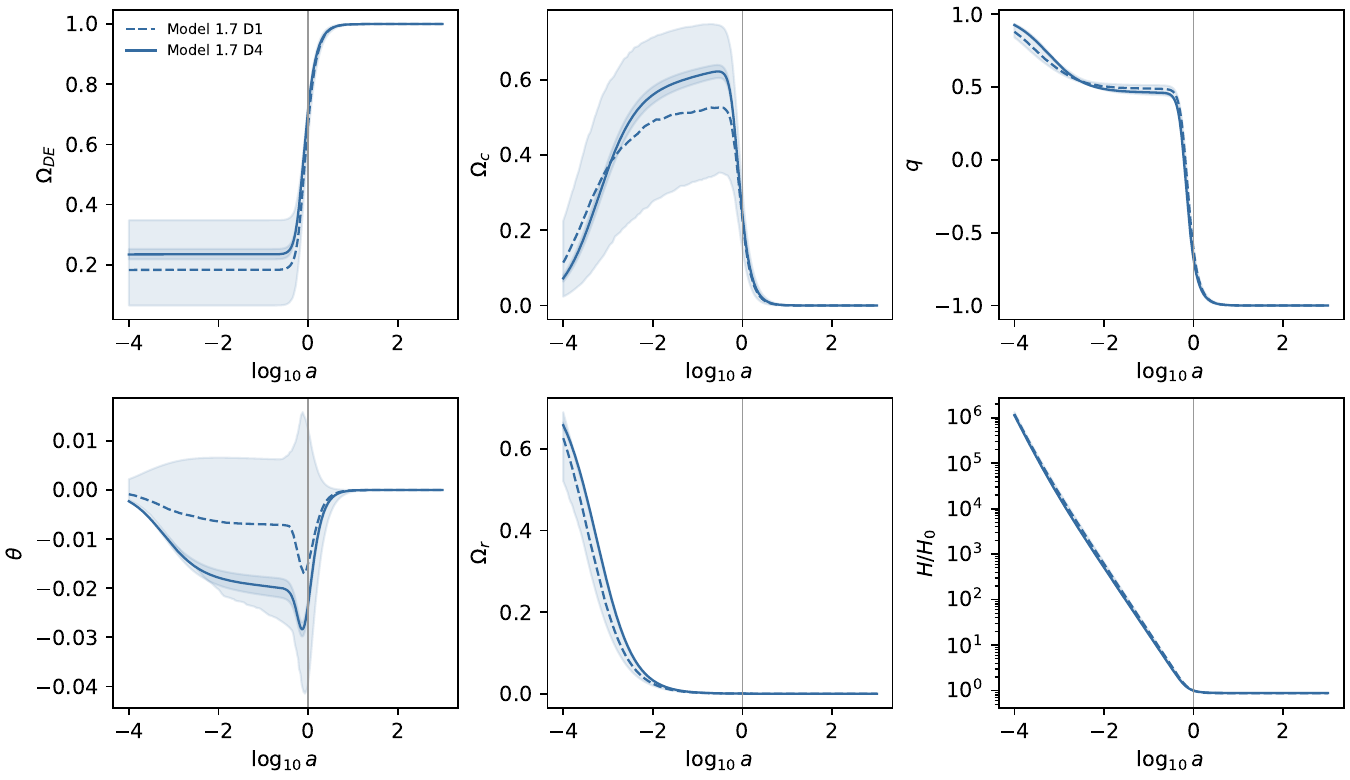}
\caption{Background histories of Model 1.7. Dashed and solid curves denote D1 and D4; bands show pointwise central $68.3\%$ ranges for 400 joint vectors per group. The horizontal coordinate is $\log_{10}a$, with today at zero. Negative $\theta$ denotes DE-to-DM transfer. At the DE-dominated endpoint, the product interaction vanishes.}\label{fig:model17_histories}
\end{figure}

\begin{table}[!htbp]\centering\small
\caption{Acceleration-transition redshifts from the 400 histories in each group. All selected histories reach $a=10^3$.}\label{tab:cpc_trajectories}
\begin{tabular}{lcc}\toprule
Model & D1 & D4\\\midrule
HHDE & $0.518^{+0.038}_{-0.036}$ & $0.787^{+0.008}_{-0.010}$\\
Model 1.7 & $0.530^{+0.043}_{-0.041}$ & $0.657^{+0.018}_{-0.020}$\\
Model 5.6 & $0.573^{+0.070}_{-0.060}$ & $0.651^{+0.018}_{-0.019}$\\
\bottomrule\end{tabular}
\end{table}

For Model 1.7, the acceleration transition $q(a_t)=0$ satisfies
\begin{equation}
5x(a_t)-3\eta x(a_t)y(a_t)=1+\Omega_r(a_t)+4\alpha .\label{eq:17acceleration}
\end{equation}
The coupling enters both this relation and the evolution of the dark densities. The D4 trajectories have a transition redshift $z_t=0.657^{+0.018}_{-0.020}$ and approach $(x_*,y_*,z_*)=(1,0,0)$ smoothly. DE-to-DM transfer slows the dilution of DM as its fraction approaches zero.

For Model 5.6, $q=0$ gives $Q=0$ and the transition condition
\begin{equation}
5x(a_t)+9\xi_{DM}y(a_t)=1+\Omega_r(a_t)+4\alpha .\label{eq:acceleration_condition}
\end{equation}
At $\xi_{DM}=0$, this condition reduces to the HHDE transition condition. The coupling affects $a_t$ through the density histories. For $\gamma>0$, energy flows from DM to DE during deceleration and reverses during acceleration. At the endpoint, $\gamma=(1-3\xi_{DM})y_*$ balances dilution against exchange and viscous stress, allowing a finite residual DM fraction in de~Sitter expansion.

\begin{figure}[!htbp]\centering
\includegraphics[width=\linewidth]{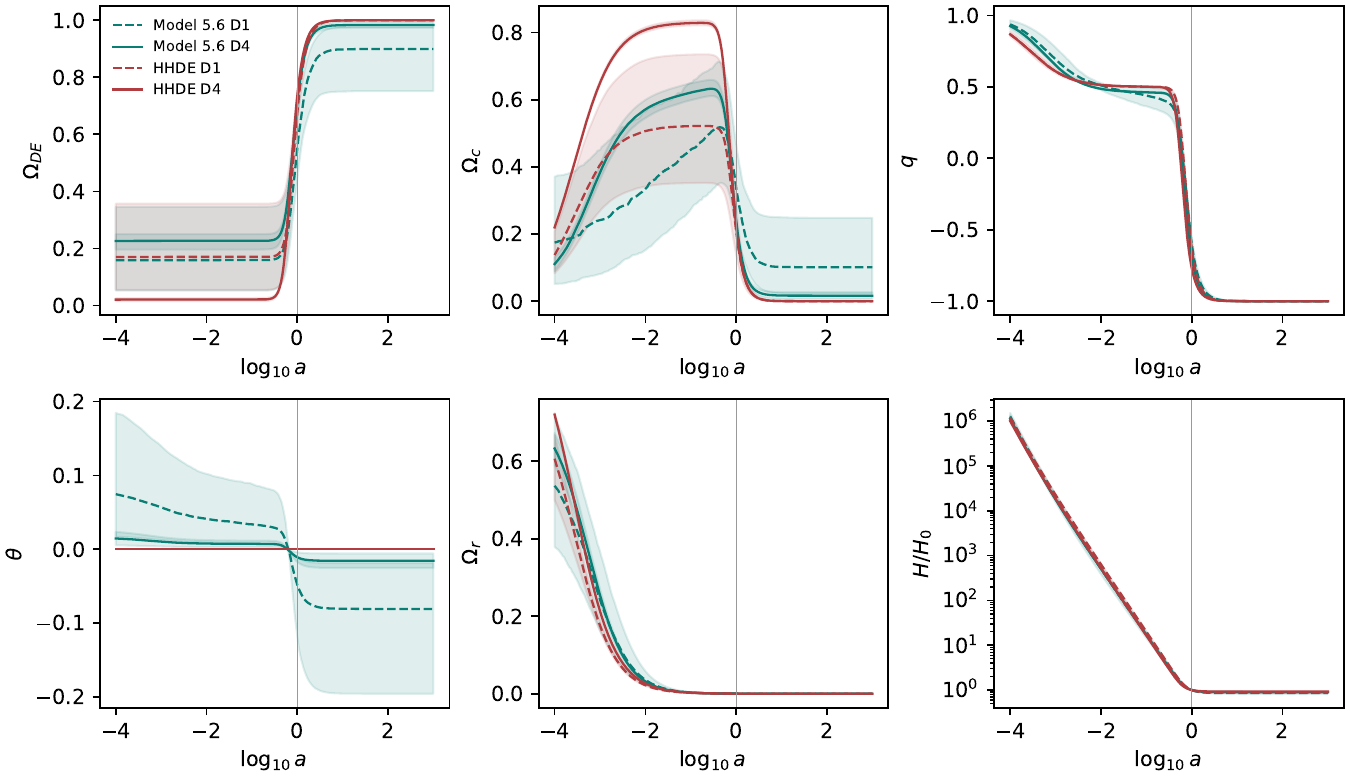}
\caption{HHDE and Model-5.6 histories, with the same conventions as Figure~\ref{fig:model17_histories}. HHDE has $\theta=0$.}\label{fig:cpc_trajectories}
\end{figure}

Models 1.7 and 5.6 have similar D4 acceleration epochs and different asymptotic compositions: $y_*=0$ in Model 1.7 and $y_*=\gamma/(1-3\xi_{DM})$ in Model 5.6.

\subsection{Sound-horizon calibration and Hubble-constant profiles}\label{sec5.calibration}

We combine Pantheon, the 31 CC measurements, and DESI DR2 and treat the drag-epoch sound horizon $r_d$ as an independent positive length to separate the background distance response from early-time acoustic calibration. This sample contains 1092 observables and excludes both the Planck distance prior and a recombination-based prediction for $r_d$.

For the free-$r_d$ profiles, we restrict Model 5.6 to $\gamma\in[0,0.3]$. At fixed $H_0$, we optimize the remaining background parameters within the ranges of Section~\ref{sec5.data}, requiring nonnegative densities and the regular positive-$\beta$ branch for HHDE and both extensions. We profile the supernova nuisance parameter $M$ over $[-25,-15]$. With $s_d=1/r_d$, the BAO contribution is
\begin{equation}
\chi^2_{\rm BAO}(s_d)
=(s_d\boldsymbol B-\boldsymbol d)^T C_{\rm BAO}^{-1}
 (s_d\boldsymbol B-\boldsymbol d),
\label{eq:freerdchi}
\end{equation}
where $\boldsymbol B$ contains the predicted $D_M$, $D_H$, or $D_V$ distances in the order of the DESI measurements $\boldsymbol d$. At cosmological redshift $z_{\rm obs}$, distinct from the phase-space coordinate $z$, these distances are
\begin{align*}
D_M(z_{\rm obs})&=c_{\rm light}\int_0^{z_{\rm obs}}\frac{d\zeta}{H(\zeta)},
\\ D_H(z_{\rm obs})&=\frac{c_{\rm light}}{H(z_{\rm obs})},\\
D_V(z_{\rm obs})&=[z_{\rm obs}D_M^2(z_{\rm obs})D_H(z_{\rm obs})]^{1/3},
\end{align*}
where $c_{\rm light}$ is the speed of light. We retain the full released covariance.

The conditional minimum is
\begin{equation*}
\widehat s_d=\frac{\boldsymbol B^T C_{\rm BAO}^{-1}\boldsymbol d}{\boldsymbol B^T C_{\rm BAO}^{-1}\boldsymbol B},
\end{equation*}
restricted to the interval corresponding to $60\leq r_d/\mathrm{Mpc}\leq250$. All resulting optima lie inside the sound-horizon boundaries.

We evaluate the profiles at $H_0=64,65,\ldots,78\,\mathrm{km\,s^{-1}\,Mpc^{-1}}$ for $\Lambda$CDM, HHDE, Model 1.7, and Model 5.6, with and without the independent-$r_d$ BAO sample. At each of the 120 points, at least two optimization starts give $\chi^2$ values that agree within $0.05$. Tightening the integration tolerance tenfold changes $\chi^2$ by less than $1.1\times10^{-7}$.

\begin{figure}[!htbp]
\centering
\includegraphics[width=\textwidth]{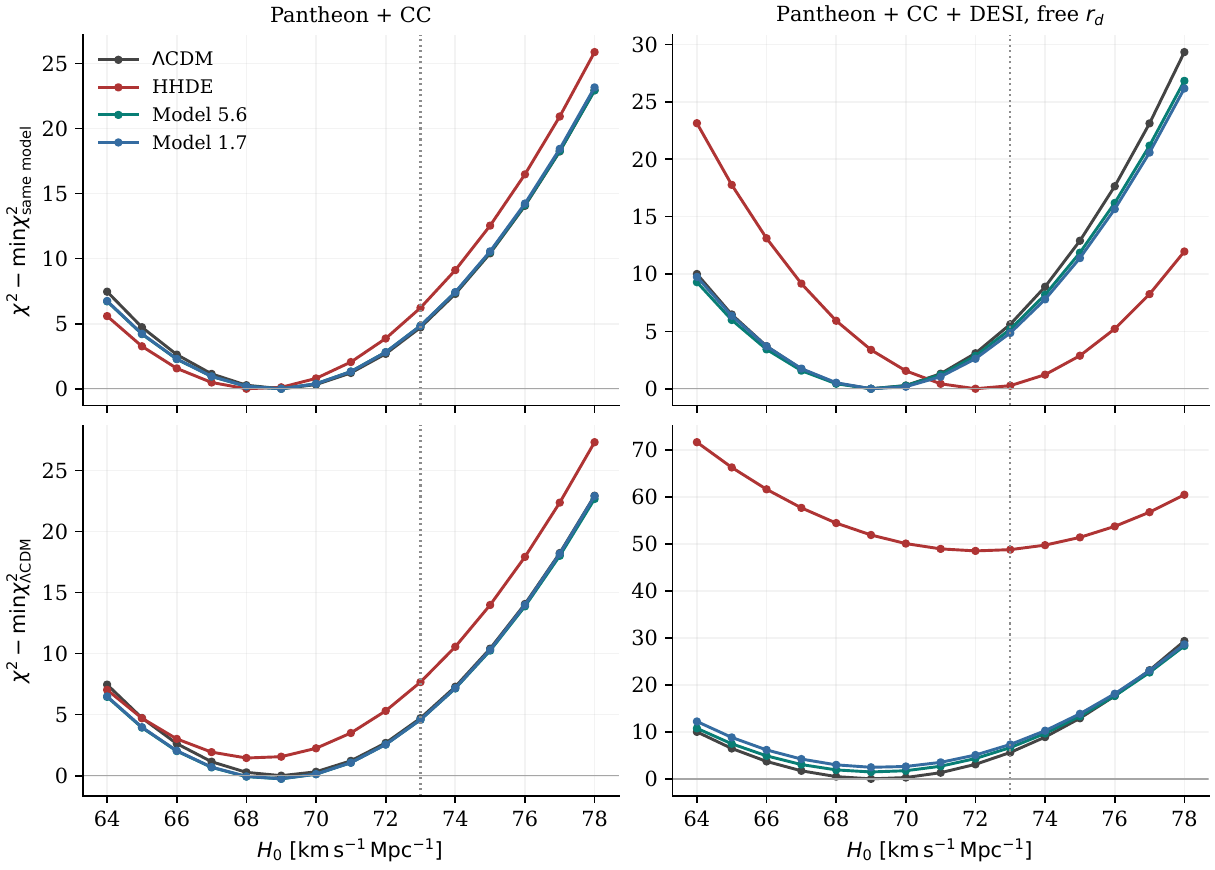}
\caption{Hubble-constant likelihood profiles with the remaining parameters optimized within the stated domains. Left: Pantheon+CC. Right: the same data plus DESI DR2 with independent $r_d$, without Planck. The upper row subtracts each model's own grid minimum; the lower row uses the $\Lambda$CDM grid minimum for the same dataset. Lines connect evaluated points, and the vertical dotted line marks $H_0=73\,\mathrm{km\,s^{-1}\,Mpc^{-1}}$.}
\label{fig:cpc_profiles}
\end{figure}

For each model and dataset, the upper row of Figure~\ref{fig:cpc_profiles} shows
\begin{equation*}
\Delta\chi^2_{\rm prof}(H_0)=\chi^2_{\rm prof}(H_0)-\min_{\rm grid}\chi^2_{\rm prof}.
\end{equation*}
Table~\ref{tab:cpc_profiles} reports the grid minima and the relative fit cost at $H_0=73\,\mathrm{km\,s^{-1}\,Mpc^{-1}}$. The grid spacing sets the resolution of the minimizing $H_0$ values.

\begin{table}[!htbp]
\centering\small
\caption{Results on the profile grid. $H_{0,g}$ is the minimizing grid value in km s$^{-1}$ Mpc$^{-1}$, $r_{d,g}$ is its conditional sound horizon in Mpc, and $\Delta\chi^2_{73}$ is evaluated relative to that model's grid minimum.}
\label{tab:cpc_profiles}
\begin{tabular}{llrrrr}\toprule
Model & Data & $H_{0,g}$ & $\chi^2_g$ & $r_{d,g}$ & $\Delta\chi^2_{73}$\\\midrule
$\Lambda$CDM & D1 & 69 & 1040.23 & -- & 4.71 \\
HHDE & D1 & 68 & 1041.68 & -- & 6.22 \\
Model 5.6 & D1 & 69 & 1039.97 & -- & 4.84 \\
Model 1.7 & D1 & 69 & 1039.98 & -- & 4.84 \\
$\Lambda$CDM & D1+BAO ($r_d$ free) & 69 & 1050.56 & 147.18 & 5.62 \\
HHDE & D1+BAO ($r_d$ free) & 72 & 1099.06 & 147.13 & 0.27 \\
Model 5.6 & D1+BAO ($r_d$ free) & 69 & 1052.02 & 147.17 & 5.19 \\
Model 1.7 & D1+BAO ($r_d$ free) & 69 & 1053.01 & 147.41 & 4.86 \\
\bottomrule\end{tabular}

\end{table}

Pantheon+CC places all four minima near $68$--$69\,\mathrm{km\,s^{-1}\,Mpc^{-1}}$. With independently calibrated BAO, the HHDE minimum moves to 72 and the relative cost at 73 falls from $6.22$ to $0.27$. The HHDE high-$H_0$ branch therefore persists when the sound horizon is independently fitted and the Planck distance prior is omitted.

In the free-$r_d$ sample, the HHDE grid minimum exceeds that of $\Lambda$CDM by $48.50$ in $\chi^2$. Model 5.6 reduces this difference to $1.47$, with its minimum near 69 and a relative cost of $5.19$ at 73.

The conditional sound horizons at the four minima lie within $147.1$--$147.4$ Mpc, well inside their allowed interval. For this free-$r_d$ combination, HHDE minimizes at $\alpha=0.001$, and Model 5.6 approaches $\alpha=0.5$, $\xi_{DM}=0.1$, and $\gamma=0$. The minima lie at or near parameter boundaries, so the comparison depends on the stated parameter domain.

For Model 1.7, the independent-$r_d$ grid minimum is $\chi^2_g=1053.01$ at $H_{0,g}=69$ km s$^{-1}$ Mpc$^{-1}$, $2.46$ above the $\Lambda$CDM minimum. Its cost at 73 is $\Delta\chi^2_{73}=4.86$, and $r_{d,g}=147.41$ Mpc. The minimum approaches $\alpha=0.5$ and $\eta=-0.3$. The difference between this minimum and the D4 result shows that the inferred $H_0$ in Model 1.7 depends on acoustic calibration.

\subsection{Fit quality and the Hubble-constant discrepancy}\label{sec5.comparison}

For each common data combination, we report the Akaike and Bayesian information criteria~\cite{Akaike1974,Schwarz1978},
\begin{equation}
\mathrm{AIC}=\chi^2_{\min}+2k,\qquad \mathrm{BIC}=\chi^2_{\min}+k\ln N_{\rm data},\label{eq:information}
\end{equation}
where $k$ includes $M$ and every independently sampled parameter. Table~\ref{tab:comparison} uses $\Lambda$CDM as the reference within each dataset. Parameter-independent likelihood normalizations cancel in these differences. Weakly identified directions and boundary optima limit the usual asymptotic interpretation of the information-criterion differences.

\begin{table}[!htbp]\centering\small
\caption{Numerical likelihood maxima and information-criterion differences within the stated parameter domains. The dagger refers to the associated sample diagnostic in Table~\ref{tab:fit1}.}\label{tab:comparison}
\begin{adjustbox}{max width=\linewidth}\begin{tabular}{llrrrr}
\toprule
Model & Data & $k$ & $\chi^2_{\min}$ & $\Delta\mathrm{AIC}$ & $\Delta\mathrm{BIC}$\\
\midrule
$\Lambda$CDM & D1 & 5 & 1040.23 & 0.00 & 0.00\\
$w$CDM & D1 & 6 & 1040.04 & 1.81 & 6.80\\
HHDE & D1 & 6 & 1041.65 & 3.42 & 8.40\\
Model 1.7 & D1 & 7 & 1039.97 & 3.74 & 13.71\\
Model 5.6 & D1 & 8 & 1039.98 & 5.75 & 20.70\\
$\Lambda$CDM & D4 & 5 & 1055.02 & 0.00 & 0.00\\
$w$CDM & D4 & 6 & 1054.69 & 1.68 & 6.67\\
HHDE & D4 & 6 & 1156.89 & 103.87 & 108.87\\
Model 1.7 & D4 & 7 & 1081.73 & 30.71 & 40.71\\
Model 5.6$^{\dagger}$ & D4 & 8 & 1081.87 & 32.85 & 47.84\\
\bottomrule
\end{tabular}
\end{adjustbox}
\end{table}

D1 gives similar maximum likelihoods across the models. Under D4, HHDE has $\chi^2_{\min}=1156.89$, compared with 1055.02 for $\Lambda$CDM. The information criteria add parameter penalties to the chi-squared difference of 101.87. The Pantheon and DESI contributions increase by approximately 35.77 and 66.62, respectively; the reduction in the Planck-prior contribution is much smaller. The joint conditional likelihood disfavours the specified HHDE background, which yields a higher $H_0$.

The numerical minima of Models 1.7 and 5.6 are 1081.73 and 1081.87. Additional dark-sector dynamics improve the distance history, reducing these minima relative to HHDE by 75.16 and 75.03. Model 1.7 achieves this fit with a single coupling parameter and satisfies the sampling criteria. Model 5.6 includes both coupling and viscosity and gives a nearly equal numerical optimum. Both minima exceed that of $\Lambda$CDM by approximately 27, so the information criteria favor the reference cosmology.

\begin{figure}[!htbp]\centering
\includegraphics[width=\linewidth]{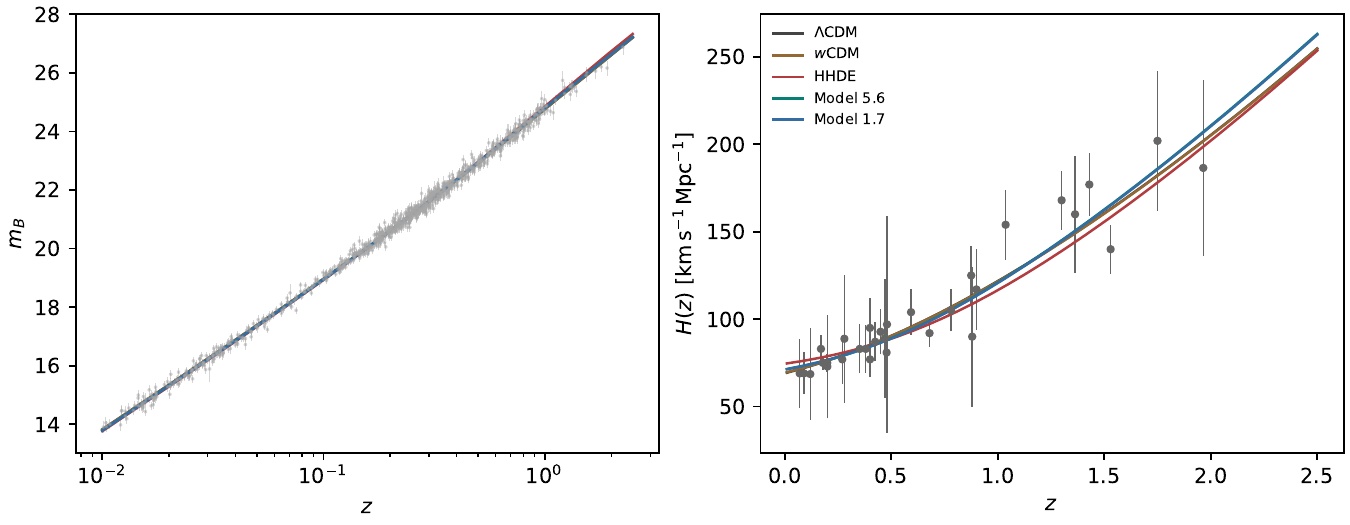}
\caption{Optimized D4 expansion histories and apparent-magnitude curves compared with Pantheon and CC measurements. Each curve uses its own optimized $M$. Supernova error bars show statistical uncertainties; the fits use the full covariance.}\label{fig:backgroundcomparison}
\end{figure}

Using the SH0ES measurement~\cite{WOS:000828873600001} as an external benchmark, a one-dimensional Gaussian comparison gives discrepancies of approximately $0.9\sigma$ for HHDE, $1.0\sigma$ for Model 1.7, and $2.2\sigma$ for D4 $\Lambda$CDM. Model 1.7 combines a reduced $H_0$ discrepancy with a substantial improvement over the HHDE distance fit. The inferred expansion scale depends on acoustic calibration, as quantified by the independent-$r_d$ profiles in Section~\ref{sec5.calibration}.

\subsection{Dark-sector parameters and physical interpretation}\label{sec5.2}

Model 1.7 has $Q=\eta\rho_{DM}\rho_{DE}/H$: transfer vanishes when either dark component vanishes. Negative $\eta$ denotes DE-to-DM transfer under Eq.~\eqref{2.9}; the exchange disappears as the DM fraction tends to zero at the de~Sitter endpoint. Model 5.6 can instead maintain a finite residual DM fraction.

\begin{figure}[!htbp]\centering
\includegraphics[width=0.98\linewidth]{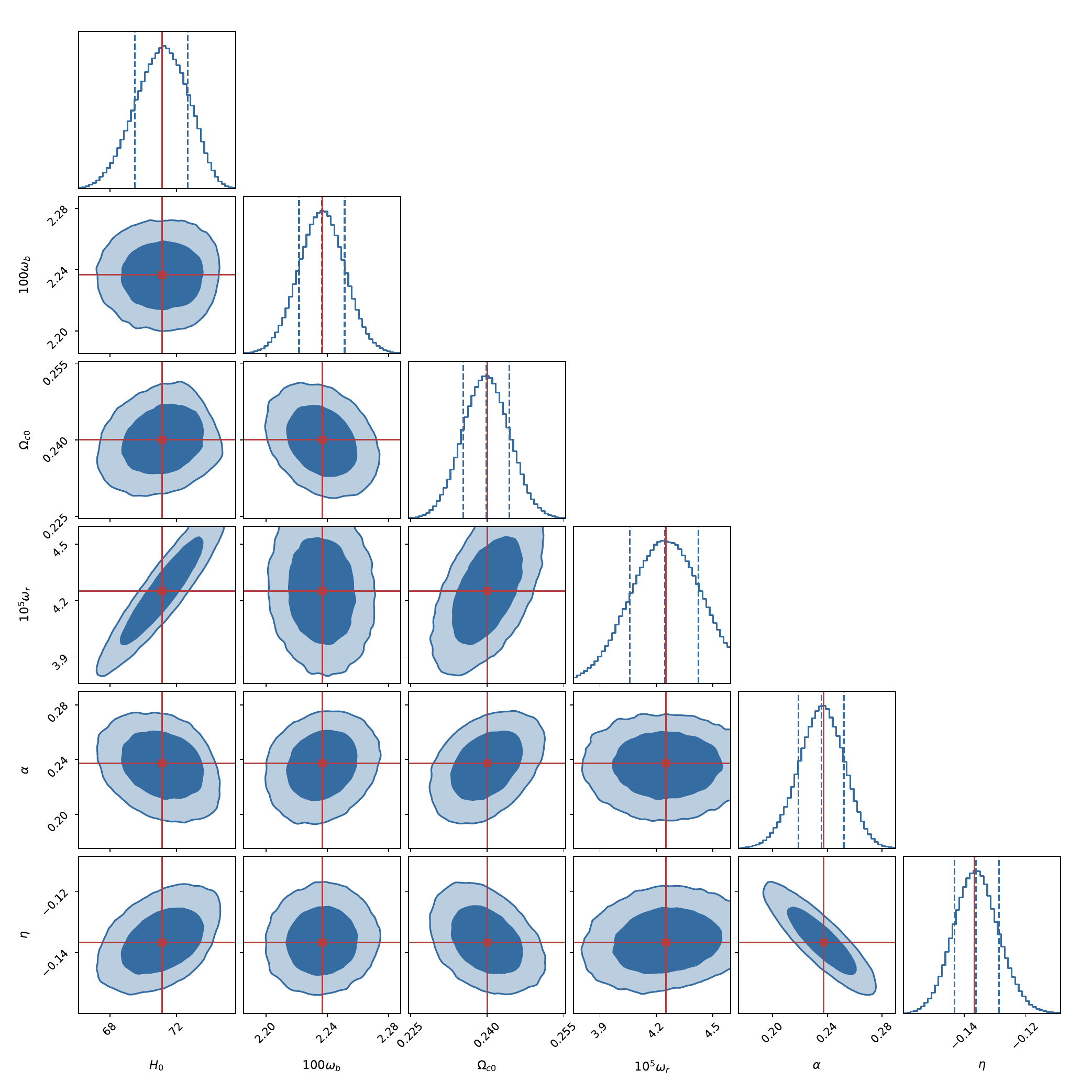}
\caption{D4 posterior distribution of Model 1.7, marginalized over $M$. Contours enclose $68.3\%$ and $95.4\%$ of the estimated density. Red lines locate the joint numerical likelihood maximum; dashed lines mark marginal quantiles. The radiation axis spans its full prior. Appendix~\ref{appen.E} gives the sampling diagnostics.}\label{fig:triangle17D4}
\end{figure}

D1 gives $\eta=-0.110^{+0.213}_{-0.137}$, with zero inside its central $68.3\%$ interval. D4 gives $\eta=-0.1361^{+0.0075}_{-0.0071}$ and selects DE-to-DM transfer within the specified background likelihood. For Model 5.6, D4 gives $\gamma=0.0144^{+0.0100}_{-0.0090}$ and $\xi_{DM}=0.0143^{+0.0029}_{-0.0025}$, with strong correlations. At the median parameter vector, these coefficients imply a residual DM fraction $y_*\simeq0.015$ at the de~Sitter endpoint.

At high redshift the HHDE density fraction approaches $\alpha$. The D4 median of this limit is approximately 0.020 in HHDE and 0.22--0.24 in the extensions. This early component affects expansion during recombination, so the distance-prior approximation requires assessment with a full CMB calculation. Such a calculation requires a covariant interaction prescription and a closure for dissipative perturbations.
\FloatBarrier

\section{Summary and outlook}\label{sec6}

The HHDE density definition and component continuity equations give a common background description of the 35 interacting and viscous fluid models. Reconstructing the intrinsic DE pressure makes Classes 2 and 3 background-degenerate with Class 1. Interaction and DM viscosity enter through their net contribution, yielding exact reparameterizations between Models 4.4 and 1.4 and between Models 5.1 and 1.1. Expansion observations therefore constrain the corresponding effective parameter combinations.

The regular stationary endpoints on the positive-$\beta$, finite-$H$ branch are de~Sitter, with analytic local existence and stability conditions for each of the seven interactions. The 2400 selected HHDE and extension trajectories connect the inferred present states to stable endpoints over the tested interval. Models 1.7 and 5.6 have similar D4 acceleration epochs but different late-time matter contents: Model 1.7 approaches a DE-dominated state, whereas Model 5.6 can maintain a finite DM fraction through exchange and viscous stress.

D1 yields $H_0\simeq68$--$69$ km s$^{-1}$ Mpc$^{-1}$ across the five cosmologies and admits zero interaction in Model 1.7. With recombination-based acoustic scales and the conditional Planck distance prior, Model 1.7 gives $H_0=71.15^{+1.53}_{-1.64}$ km s$^{-1}$ Mpc$^{-1}$ and $\eta=-0.1361^{+0.0075}_{-0.0071}$. With one additional coupling parameter, its DE-to-DM transfer lowers the minimum $\chi^2$ by 75.16 relative to HHDE. The inferred $H_0$ differs from SH0ES by approximately $1\sigma$ in a one-dimensional comparison. The information criteria favor $\Lambda$CDM, with a remaining difference of $\Delta\chi^2=26.71$ relative to this reference cosmology. The independent ensembles for Model 1.7 give consistent posterior distributions. Model 5.6 gives a comparable numerical optimum through energy exchange and dissipative pressure.

With $r_d$ fitted independently, HHDE prefers a high expansion scale and has a substantially larger minimum $\chi^2$ than the reference model. Model 1.7 substantially improves this fit and has its grid minimum near $69$ km s$^{-1}$ Mpc$^{-1}$. The difference between this result and D4 establishes that the inferred Hubble constant depends on the acoustic information used.

A perturbative extension requires a covariant prescription for energy transfer and dissipative stresses. Full CMB spectra and structure growth can then test the early dark component and assess the applicability of the compressed distance prior used here.
\FloatBarrier

\section*{Acknowledgments}
\paragraph{Generative AI disclosure.}
During manuscript preparation, the authors used DeepSeek and GPT-6 for language editing and condensation. Computational uses are described in Section~\ref{sec5.data}. The authors developed the scientific model, original codes and initial manuscript, and retain full responsibility for reviewing the AI-assisted material and for the content of this article.
\paragraph{Competing interests.}
The authors declare no competing interests.

\section*{Data and Code Availability}
The observational datasets are publicly available from the releases cited in the text. The Python background models, likelihood routines, fitting scripts, and Mathematica fixed-point calculations are available from the corresponding author upon reasonable request. Appendix~\ref{appen.D} summarizes the symbolic construction used for the dynamical analysis.

\FloatBarrier
%===========================================================
\appendix
%===========================================================

\section{Phase diagrams for Models 1.1--1.7}\label{appen.A}

\begin{figure}[!htbp]
\centering
\subfloat[Model 1.1]{\includegraphics[width=0.27\textwidth]{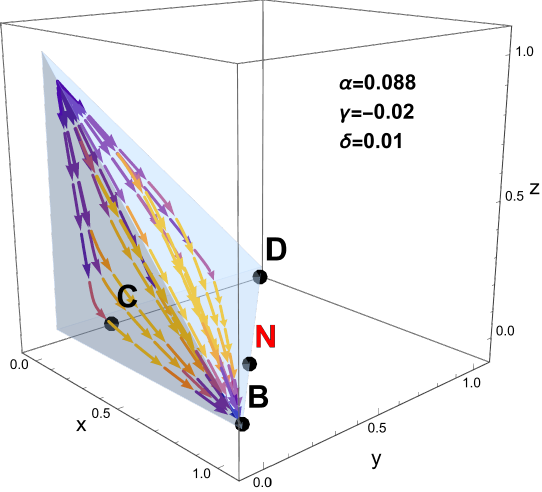}}\hfill
\subfloat[Model 1.2]{\includegraphics[width=0.27\textwidth]{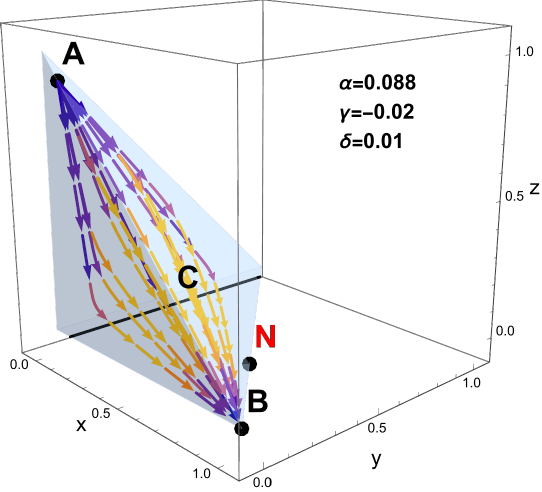}}\hfill
\subfloat[Model 1.3]{\includegraphics[width=0.27\textwidth]{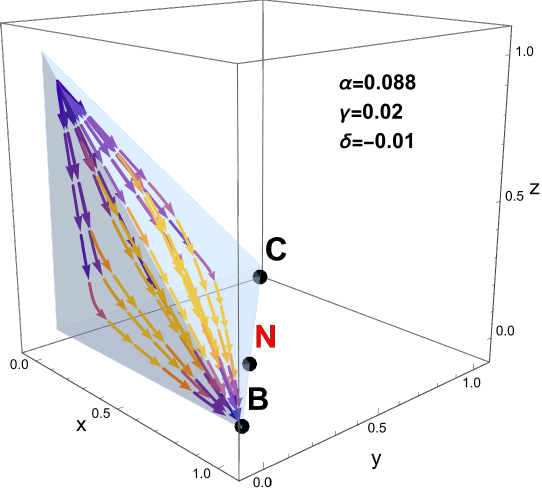}}
\vspace{0.3\baselineskip}
\subfloat[Model 1.4]{\includegraphics[width=0.27\textwidth]{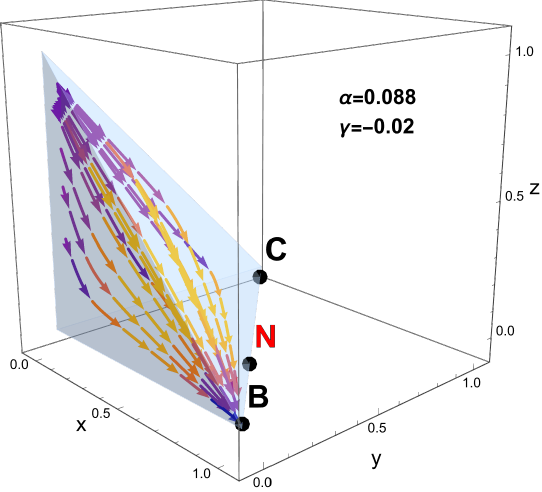}}\hfill
\subfloat[Model 1.5]{\includegraphics[width=0.27\textwidth]{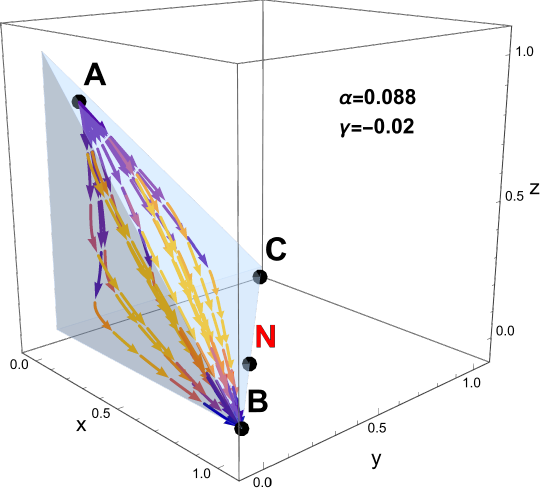}}\hfill
\subfloat[Model 1.6]{\includegraphics[width=0.27\textwidth]{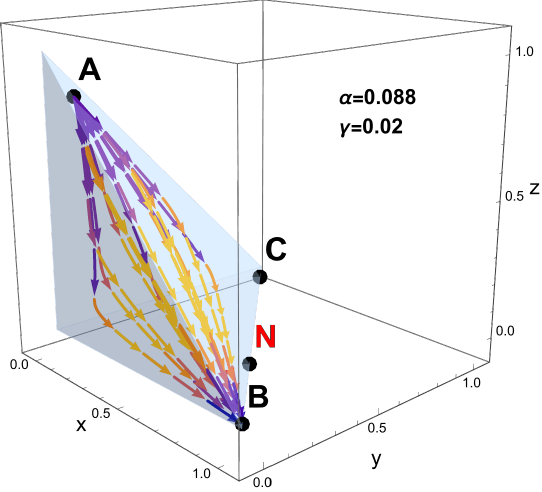}}
\vspace{0.3\baselineskip}
\subfloat[Model 1.7]{\includegraphics[width=0.27\textwidth]{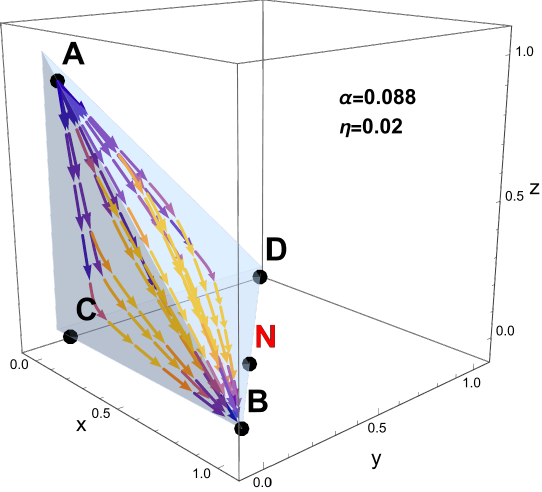}}
\caption{Phase portraits for Models 1.1--1.7. The same background portraits apply to the corresponding Models 2.x and 3.x. The red marker N denotes the fixed geometric reference $(x,y,z)=(0.683,0.317,0)$ on the $b=0$ boundary in every panel; $y$ denotes dark matter alone.}
\label{figA.1}
\end{figure}

Figure~\ref{figA.1} shows the phase portraits discussed in Section~\ref{sec4.2}, with panels (a)--(g) corresponding to Models 1.1--1.7 in order.

\FloatBarrier

\section{Complete autonomous-system specification}\label{appen.B}

We write all 35 realizations in the following common form. Let $c=1,\ldots,5$ label the viscosity class and $j=1,\ldots,7$ the interaction prescription listed in Section~\ref{sec2}. We define
\begin{equation}
\mathcal{G}_{c,j}=\frac{-1-z/3-\theta_j+x+v_{DM}^{(c)}}{1+x-2\alpha},
\qquad
\boldsymbol{F}_{c,j}=\begin{pmatrix}
-2(x-\alpha)\mathcal{G}_{c,j}\\
-y-\theta_j+v_{DM}^{(c)}-y\mathcal{G}_{c,j}\\
-4z/3-z\mathcal{G}_{c,j}
\end{pmatrix}.
\label{B.1}
\end{equation}
Eliminating the derivatives in $\theta_j$ gives an explicit rational vector field:
\begin{equation}
\theta_j=A_j+B_j\mathcal G,\qquad
\mathcal G_{c,j}=
\frac{-1-z/3+x+v-A_j}{1+x-2\alpha+B_j},
\qquad v=v_{DM}^{(c)}.
\label{B.2}
\end{equation}
Table~\ref{tab:Binteractions} lists the coefficients. The reductions for $j=2$ and $j=3$ require $1+\delta\ne0$ and $1+\gamma\ne0$, respectively; both conditions hold throughout the scanned ranges. The regular branch excludes zeros of $1+x-2\alpha+B_j$.

\begin{table}[!htbp]
\centering\small
\caption{Interaction coefficients in Eq.~\eqref{B.2}; $v$ is the DM-viscosity term for each class.}
\label{tab:Binteractions}
\begin{tabular}{ccc}
\toprule
$j$ & $A_j$ & $B_j$\\
\midrule
1 & $\delta y+\gamma x$ & $0$\\[3pt]
2 & $\dfrac{\delta(v-y)}{1+\delta}$
  & $-\dfrac{\delta y+2\gamma(x-\alpha)}{1+\delta}$\\[6pt]
3 & $\dfrac{\delta(x+y)+\gamma(v-y)}{1+\gamma}$
  & $-\dfrac{\gamma[2(x-\alpha)+y]}{1+\gamma}$\\[6pt]
4 & $\gamma$ & $0$\\[3pt]
5 & $0$ & $\gamma$\\[3pt]
6 & $-\gamma$ & $-3\gamma/2$\\[3pt]
7 & $\eta xy$ & $0$\\
\bottomrule
\end{tabular}
\end{table}

The viscous terms for each class are
\begin{table}[!htbp]
\centering
\caption{Viscous terms defining the five model classes.}
\label{tab:Bclasses}
\begin{tabular}{ccc}
\toprule
Class & $v_{DE}^{(c)}$ & $v_{DM}^{(c)}$\\
\midrule
1 & $0$ & $0$\\
2 & $3\xi_0$ & $0$\\
3 & $3\xi_{DE}x$ & $0$\\
4 & $0$ & $3\xi_1$\\
5 & $0$ & $3\xi_{DM}y$\\
\bottomrule
\end{tabular}
\end{table}
Equation~\eqref{B.1} and Table~\ref{tab:Bclasses} give
\[
\boldsymbol{F}_{2,j}=\boldsymbol{F}_{3,j}=\boldsymbol{F}_{1,j},
\]
The backgrounds of Classes 4 and 5 generally also depend on DM viscosity. The intrinsic DE equations of state of Classes 2 and 3 satisfy
\[
\omega_{DE}^{(2,j)}=\omega_{DE}^{\rm eff,(1,j)}+\frac{3\xi_0}{x},
\qquad
\omega_{DE}^{(3,j)}=\omega_{DE}^{\rm eff,(1,j)}+3\xi_{DE}.
\]

A single algebraic condition determines the finite-$H$ fixed points. Substituting the stationary relations in Eq.~\eqref{eq:regular_ds} into the numerator of Eq.~\eqref{B.2} gives $v_*-A_j(x_*,y_*)=y_*$ and the results in Table~\ref{tab:regular_ds}. In coordinates $(b,z,x)$, the linearized system is block triangular: the baryon and radiation eigenvalues are $-1$ and $-4/3$, and the derivative along $y=1-x$ gives the remaining eigenvalue in Eq.~\eqref{eq:regular_spectrum}.

\begin{table}[!htbp]
\centering\small
\caption{De~Sitter candidates on the regular positive-$\beta$ branch with finite $H$. Each has $x_*=1-y_*$ and $z_*=b_*=0$; Eq.~\eqref{eq:regular_spectrum} determines its stability.}
\label{tab:regular_ds}
\begin{tabular}{ccc}
\toprule
$j$ & $y_*$ & $\kappa_*$\\
\midrule
1 & $(\nu-\gamma)/(s+\delta-\gamma)$ & $s+\delta-\gamma$\\[3pt]
2 & $\nu/s$ & $s/(1+\delta)$\\[3pt]
3 & $(\nu-\delta)/s$ & $s/(1+\gamma)$\\[3pt]
4 & $(\nu-\gamma)/s$ & $s$\\[3pt]
5 & $\nu/s$ & $s$\\[3pt]
6 & $(\nu+\gamma)/s$ & $s$\\[3pt]
7 & $\eta y_*^2-(s+\eta)y_*+\nu=0$ & $s+\eta(1-2y_*)$\\
\bottomrule
\end{tabular}
\end{table}

The explicit fractions require nonzero denominators. For interaction 7, we test every real root of the quadratic; at $\eta=0$, it reduces to $s y_*=\nu$. The criterion excludes parameter values that invalidate the algebraic elimination, singular cases with $D_*=0$, and non-hyperbolic cases with $\kappa_*=0$. The high-$H$ boundary $x=\alpha$ is distinct from these finite-$H$ endpoints. For an accepted candidate, $H_*=H_0[\beta/(x_*-\alpha)]^{1/4}$. On the grid of Eq.~\eqref{4.3}, these analytic conditions reproduce every accepted count in Table~\ref{tab:scanconditions}.

\section{Representative DM-viscosity attractors and spectra}\label{appen.C}

Tables~\ref{tab:C4} and~\ref{tab:C5} list representative stable physical fixed points at \(\alpha=0.088\). Unlisted interaction parameters are zero. These parameter choices satisfy the grid conditions in Table~\ref{tab:scanconditions}.

\begin{table}[!htbp]
\centering
\scriptsize
\caption{Representative Class-4 attractors.}
\label{tab:C4}
\resizebox{\textwidth}{!}{%
\begin{tabular}{ccccc}
\toprule
Model & Parameters & Attractor \((x,y,z)\) & Jacobian eigenvalues & \(\omega_{DE}^{\rm eff}\)\\
\midrule
4.1 & \(\gamma=-0.01,\delta=0.04,\xi_1=0.001\) & \((0.988,0.012,0)\) & \((-1.333,-1.043,-1.000)\) & \(-1.009\)\\
4.2 & \(\gamma=-0.02,\delta=-0.01,\xi_1=0.005\) & \((0.985,0.015,0)\) & \((-1.333,-1.000,-0.982)\) & \(-1.000\)\\
4.3 & \(\gamma=0.02,\delta=0.01,\xi_1=0.005\) & \((0.995,0.005,0)\) & \((-1.333,-1.000,-0.997)\) & \(-0.990\)\\
4.4 & \(\gamma=0.02,\xi_1=0.010\) & \((0.990,0.010,0)\) & \((-1.333,-1.000,-0.994)\) & \(-0.980\)\\
4.5 & \(\gamma=0.02,\xi_1=0.001\) & \((0.997,0.003,0)\) & \((-1.333,-1.000,-0.988)\) & \(-1.000\)\\
4.6 & \(\gamma=-0.01,\xi_1=0.005\) & \((0.995,0.005,0)\) & \((-1.333,-1.000,-0.989)\) & \(-0.990\)\\
4.7 & \(\eta=0.02,\xi_1=0.005\) & \((0.985,0.015,0)\) & \((-1.333,-1.011,-1.000)\) & \(-1.000\)\\
\bottomrule
\end{tabular}}
\end{table}

\begin{table}[!htbp]
\centering
\scriptsize
\caption{Representative Class-5 attractors for \(\xi_{DM}=0.005\).}
\label{tab:C5}
\resizebox{\textwidth}{!}{%
\begin{tabular}{ccccc}
\toprule
Model & Interaction parameters & Attractor \((x,y,z)\) & Jacobian eigenvalues & \(\omega_{DE}^{\rm eff}\)\\
\midrule
5.1 & \(\gamma=0,\delta=0.01\) & \((1,0,0)\) & \((-1.333,-1.000,-0.995)\) & \(-1.000\)\\
5.2 & \(\gamma=0.02,\delta=0.01\) & \((1,0,0)\) & \((-1.333,-1.000,-0.995)\) & \(-1.000\)\\
5.3 & \(\gamma=-0.02,\delta=0\) & \((1,0,0)\) & \((-1.333,-1.000,-0.985)\) & \(-1.000\)\\
5.4 & \(\gamma=-0.02\) & \((0.980,0.020,0)\) & \((-1.333,-1.000,-0.974)\) & \(-1.020\)\\
5.5 & \(\gamma=0.02\) & \((1,0,0)\) & \((-1.333,-1.000,-0.974)\) & \(-1.000\)\\
5.6 & \(\gamma=0.02\) & \((0.980,0.020,0)\) & \((-1.333,-1.000,-0.990)\) & \(-1.020\)\\
5.7 & \(\eta=0.02\) & \((1,0,0)\) & \((-1.333,-1.005,-1.000)\) & \(-1.000\)\\
\bottomrule
\end{tabular}}
\end{table}
All attractors in Tables~\ref{tab:C4} and~\ref{tab:C5} have \(q=-1\).

\section{Mathematica implementation}\label{appen.D}

The Mathematica code below constructs the autonomous systems and scans their critical points. Here \(g=2H'/H\), \texttt{th} denotes \(\theta\), and \texttt{c} and \texttt{i} label the viscosity class and interaction, respectively.
\begin{verbatim}
vars = {x, y, z};
vDE = {0, 3 xi0, 3 xiDE x, 0, 0};
vDM = {0, 0, 0, 3 xi1, 3 xiDM y};
interaction = {delta y + gamma x, delta yp + gamma xp,
  delta (x + y) + gamma (xp + yp), gamma, gamma g,
  gamma (-1 - 3 g/2), eta x y};

rule[c_, i_] := rule[c, i] = First@FullSimplify@Solve[{
  g == (-1-z/3-th+x+vDM[[c]])/(1+x-2 alpha),
  xp == -2 (x-alpha) g, yp == -y-th+vDM[[c]]-y g,
  zp == -4 z/3-z g, th == interaction[[i]]},
  {g, xp, yp, zp, th}, Reals];
system[c_, i_] := system[c, i] =
  FullSimplify[Cancel[Together[#]] & /@
    ({xp, yp, zp} /. rule[c, i])];
gFun[c_, i_] := g /. rule[c, i];
jac[c_, i_] := D[system[c, i], {vars}];

physical = 0 <= x <= 1 && 0 <= y <= 1 && 0 <= z <= 1 &&
  0 <= x + y + z <= 1;
aGrid = Sort@DeleteDuplicates@Join[{33/1000},
  Range[38/1000, 138/1000, 1/100]];
cGrid = Range[-1/10, 1/10, 1/100];
vGrid = Sort@DeleteDuplicates@Join[{0},
  Range[1/200, 19/200, 1/100], {1/10}];
pNames[c_, i_] := Join[{alpha},
  Switch[i, 1|2|3, {delta, gamma}, 4|5|6, {gamma}, 7, {eta}],
  Switch[c, 4, {xi1}, 5, {xiDM}, _, {}]];
pRanges[c_, i_] := Join[{aGrid},
  Switch[i, 1|2|3, {cGrid, cGrid}, 4|5|6, {cGrid}, 7, {cGrid}],
  Switch[c, 4|5, {vGrid}, _, {}]];

zeroRules[c_, i_, par_] := Thread[
  Select[Rest[pNames[c, i]],
    TrueQ[(# /. par) == 0] &] -> 0];
factors[p_] := If[TrueQ[p === 0], {0},
  First /@ Rest[FactorList[p]]];
templates[c_, i_, zr_] := templates[c, i, zr] = Module[{f, b},
  f = factors /@ Numerator /@ Together /@ (system[c, i] /. zr);
  b = Tuples[f];
  DeleteDuplicates@Flatten[
    Quiet[(Solve[Thread[# == 0], vars] &) /@ b], 1]];
fixed[c_, i_, par_] := Module[{r, den},
  r = Quiet@FullSimplify[
    templates[c, i, zeroRules[c, i, par]] /. par];
  den = Denominator /@ Together /@ system[c, i];
  Select[r, FreeQ[vars /. #, Alternatives @@ vars] &&
    VectorQ[N[vars /. #], NumericQ] &&
    TrueQ[FullSimplify[physical /. #]] &&
    TrueQ[FullSimplify[
      And @@ Thread[(den /. par /. #) != 0]]] &]];
attractorQ[c_, i_, par_] :=
  AnyTrue[fixed[c, i, par], Function[r,
    With[{e = Quiet@Check[
        N[Eigenvalues[jac[c, i] /. par /. r]], $Failed]},
      e =!= $Failed && VectorQ[e, NumericQ] &&
        Max[Re[e]] < -10^-8]]];

scan[c_, i_] := scan[c, i] = Module[{n, grid, good},
  n = pNames[c, i]; grid = Tuples[pRanges[c, i]];
  good = Select[Thread[n -> #] & /@ grid,
    attractorQ[c, i, #] &];
  {ToString[c] <> "." <> ToString[i],
   Length[grid], Length[good],
   "representative background scan", good}];
scan[c : 2|3, i_] := Module[{base = scan[1, i]},
  {ToString[c] <> "." <> ToString[i], base[[2]], base[[3]],
   "background-degenerate with Class 1", base[[5]]}];
allScan = Table[scan[c, i], {c, 1, 5}, {i, 1, 7}];
stream = OpenWrite["MHHVIDE_35_scan.res",
  CharacterEncoding -> "UTF-8"];
Write[stream, allScan]; Close[stream];
\end{verbatim}
The symbolic system satisfies the HHDE identity, the Friedmann constraint, and the DE continuity equation. The DE-viscosity array contributes to the intrinsic pressure and cancels from the background vector field. The associated numerical data give the accepted parameter sets and classify them by interval and orthant.

\FloatBarrier
\section{Sampling diagnostics and complete parameter summaries}\label{appen.E}

We assess sampling with rank-normalized and folded split statistics, effective sample sizes (ESS), and autocorrelation estimates~\cite{Vehtari2021}. The D1 reference samples contain 44, 36, 36, and 44 walkers for $\Lambda$CDM, $w$CDM, HHDE, and Model 5.6, respectively. Their lengths and discarded segments are $(10000,3000)$, $(20000,4000)$, $(20000,4000)$, and $(40000,4000)$. Their maximum walker-split $\widehat R$ values are 1.016, 1.013, 1.013, and 1.013; minimum bulk ESS values are 3335, 3082, 3435, and 3394. These samples satisfy the reference criteria $\widehat R\leq1.03$, ESS $\geq1500$, and autocorrelation coverage $\geq50$, which allow larger sampling errors than the criteria for the recombination-based fits below. Model-1.7 D1 uses two independent 40-walker ensembles of 75000 steps, discarding 10000, with $\widehat R_i=1.0089$, bulk ESS 9815, and tail ESS 21870. Walkers within an ensemble are coupled; separately seeded ensembles provide independent replicates.

The recombination-based samples use two independently seeded ensembles, each with 40 walkers. We exclude the separate warm-up samples and the first 10000 production steps. The primary criteria require within-ensemble and selected independent-pair $\widehat R<1.01$, bulk ESS $\geq2000$, tail ESS $\geq1000$, and autocorrelation coverage $\geq100$. They also require half-chain median drift and between-ensemble quantile differences below $0.1$ posterior standard deviations, quantile Monte Carlo error below $0.03$ standard deviations, autocorrelation-time change below $5\%$, and acceptance fractions between 0.15 and 0.65. All criteria must hold at two evaluations separated by 10000 steps.

\begin{table}[!htbp]\centering\small
\caption{Sampling diagnostics for the recombination-based fits. $n$ is the production length per walker before the 10000-step discard. $\widehat R_i$ is the primary statistic for the selected independent pairs; $\widehat R_{40}$ is the supplementary maximum over all 40 fixed pairs, including derived $\Omega_{M0}$ and $\beta$ where applicable. ESS values are minima over parameters and ensembles. Daggers mark supplementary $\widehat R_{40}\geq1.01$.}\label{tab:diagnostics}
\begin{adjustbox}{max width=\linewidth}\begin{tabular}{llrrrrr}
\toprule
Model & Data & $n$ & $\widehat R_i$ & $\widehat R_{40}$ & $\mathrm{ESS}_{\rm bulk}$ & $\mathrm{ESS}_{\rm tail}$\\
\midrule
$\Lambda$CDM & D4 & 60000 & 1.00364 & 1.00902 & 24030 & 58855\\
$w$CDM & D4 & 60000 & 1.00439 & 1.00964 & 19546 & 45154\\
HHDE & D4 & 70000 & 1.00469 & 1.00600 & 24548 & 53993\\
Model 5.6$^{\dagger}$ & D4 & 210000 & 1.00941 & 1.01134 & 14374 & 16707\\
HHDE & D2 & 70000 & 1.00620 & 1.00839 & 16146 & 34879\\
HHDE$^{\dagger}$ & D3 & 60000 & 1.00789 & 1.01060 & 18434 & 40519\\
Model 1.7 & D4 & 80000 & 1.00389 & 1.00751 & 18050 & 38020\\
\bottomrule
\end{tabular}
\end{adjustbox}
\end{table}

All recombination-based samples pass the primary criterion. For HHDE D3, the supplementary statistic in the radiation direction is $\widehat R_{40}=1.01060$, slightly above its threshold. Model-1.7 D4 passes successive evaluations at 70000 and 80000 steps per ensemble. At 80000 steps, $\widehat R_i=1.00389$, $\widehat R_{40}=1.00751$, and the minimum bulk and tail ESS values are 18050 and 38020. The autocorrelation-time change is $1.77\%$, and the maximum quantile Monte Carlo error is 0.00726 posterior standard deviations. After discarding 10000 steps, the two ensembles contain 5.6 million stored draws.

Model-5.6 D4 passes successive primary evaluations at 200000 and 210000 steps per ensemble. At 210000 steps, $\widehat R_i=1.00941$, with minimum bulk and tail ESS values of 14374 and 16707. The autocorrelation-time change is $1.32\%$, and the maximum quantile Monte Carlo error is 0.00853 posterior standard deviations. After the discard, the two ensembles contain 16 million stored draws. One of the 40 fixed pairs remains marginal in $\alpha$, $\xi_{DM}$, and derived $\beta$, with $\widehat R_{40}=1.01134$. 

Tables~\ref{tab:completeD1}--\ref{tab:mleD4} report marginal summaries and jointly optimized parameter vectors. We obtain the two D4 extension optima by bounded maximization from four starting points. Doubling the background-grid density and tightening the ODE tolerances tenfold changes their $\chi^2$ by less than $8\times10^{-8}$ at the quoted optima. At each optimum and 16 retained draws, the same finite-point accuracy test gives maximum changes of $3.6\times10^{-6}$ and $4.1\times10^{-6}$ for Models 1.7 and 5.6, respectively.

\begin{table}[!htbp]\centering\small
\caption{Complete D1 medians and central $68.3\%$ intervals. $H_0$ is in km s$^{-1}$ Mpc$^{-1}$.}\label{tab:completeD1}
\begin{adjustbox}{max width=\linewidth}\begin{tabular}{lccccc}
\toprule
Parameter & $\Lambda$CDM & $w$CDM & HHDE & Model 1.7 & Model 5.6\\
\midrule
$M$ & $-19.383^{+0.055}_{-0.055}$ & $-19.395^{+0.056}_{-0.057}$ & $-19.416^{+0.053}_{-0.055}$ & $-19.413^{+0.055}_{-0.055}$ & $-19.398^{+0.055}_{-0.056}$\\
$H_0$ & $68.93^{+1.91}_{-1.84}$ & $68.71^{+1.86}_{-1.82}$ & $68.31^{+1.83}_{-1.83}$ & $68.38^{+1.86}_{-1.83}$ & $68.70^{+1.86}_{-1.84}$\\
$100\omega_b$ & $4.949^{+3.082}_{-3.051}$ & $4.986^{+3.157}_{-3.107}$ & $4.395^{+3.279}_{-2.672}$ & $4.557^{+3.219}_{-2.777}$ & $4.641^{+3.250}_{-2.760}$\\
$\Omega_{c0}$ & $0.1977^{+0.0643}_{-0.0644}$ & $0.2167^{+0.0740}_{-0.0691}$ & $0.2117^{+0.0808}_{-0.0670}$ & $0.2202^{+0.0844}_{-0.0723}$ & $0.3403^{+0.1266}_{-0.1168}$\\
$10^5\omega_r$ & $4.175^{+0.287}_{-0.282}$ & $4.180^{+0.286}_{-0.287}$ & $4.175^{+0.290}_{-0.287}$ & $4.177^{+0.285}_{-0.287}$ & $4.185^{+0.278}_{-0.286}$\\
$\Omega_{M0}$ & $0.3028^{+0.0204}_{-0.0197}$ & $0.3274^{+0.0418}_{-0.0435}$ & $0.3226^{+0.0506}_{-0.0671}$ & $0.3292^{+0.0571}_{-0.0621}$ & $0.4460^{+0.1025}_{-0.1003}$\\
$\alpha$ & -- & -- & $0.1794^{+0.1719}_{-0.1255}$ & $0.1907^{+0.1739}_{-0.1330}$ & $0.1653^{+0.1690}_{-0.1175}$\\
$\beta$ & -- & -- & $0.4954^{+0.0777}_{-0.1046}$ & $0.4730^{+0.0931}_{-0.1150}$ & $0.3667^{+0.0886}_{-0.0903}$\\
$w$ & -- & $-1.0842^{+0.1320}_{-0.1500}$ & -- & -- & --\\
$\gamma$ & -- & -- & -- & -- & $0.0843^{+0.0988}_{-0.0603}$\\
$\xi_{DM}$ & -- & -- & -- & -- & $0.0675^{+0.0225}_{-0.0313}$\\
$\eta$ & -- & -- & -- & $-0.1097^{+0.2133}_{-0.1368}$ & --\\
\bottomrule
\end{tabular}
\end{adjustbox}
\end{table}
\begin{table}[!htbp]\centering\small
\caption{Complete D4 marginal summaries. $\Omega_{M0}$ and $\beta$ are derived jointly. The dagger denotes a borderline supplementary pairwise diagnostic.}\label{tab:completeD4}
\begin{adjustbox}{max width=\linewidth}\begin{tabular}{lccccc}
\toprule
Parameter & $\Lambda$CDM & $w$CDM & HHDE & Model 1.7 & Model 5.6$^{\dagger}$\\
\midrule
$M$ & $-19.380^{+0.046}_{-0.049}$ & $-19.374^{+0.048}_{-0.050}$ & $-19.273^{+0.048}_{-0.046}$ & $-19.350^{+0.046}_{-0.050}$ & $-19.342^{+0.047}_{-0.048}$\\
$H_0$ & $69.07^{+1.48}_{-1.54}$ & $69.38^{+1.60}_{-1.65}$ & $74.70^{+1.65}_{-1.56}$ & $71.15^{+1.53}_{-1.64}$ & $71.41^{+1.59}_{-1.58}$\\
$100\omega_b$ & $2.253^{+0.012}_{-0.012}$ & $2.251^{+0.013}_{-0.013}$ & $2.230^{+0.015}_{-0.015}$ & $2.237^{+0.015}_{-0.015}$ & $2.236^{+0.015}_{-0.015}$\\
$\Omega_{c0}$ & $0.2537^{+0.0042}_{-0.0042}$ & $0.2517^{+0.0053}_{-0.0053}$ & $0.2230^{+0.0036}_{-0.0035}$ & $0.2398^{+0.0045}_{-0.0045}$ & $0.2403^{+0.0046}_{-0.0045}$\\
$10^5\omega_r$ & $4.233^{+0.176}_{-0.181}$ & $4.227^{+0.178}_{-0.181}$ & $4.074^{+0.177}_{-0.164}$ & $4.245^{+0.179}_{-0.187}$ & $4.115^{+0.275}_{-0.244}$\\
$\Omega_{M0}$ & $0.3011^{+0.0038}_{-0.0038}$ & $0.2985^{+0.0057}_{-0.0056}$ & $0.2630^{+0.0033}_{-0.0032}$ & $0.2840^{+0.0046}_{-0.0045}$ & $0.2842^{+0.0046}_{-0.0045}$\\
$\alpha$ & -- & -- & $0.0204^{+0.0057}_{-0.0057}$ & $0.2358^{+0.0163}_{-0.0171}$ & $0.2213^{+0.0264}_{-0.0258}$\\
$\beta$ & -- & -- & $0.7165^{+0.0064}_{-0.0065}$ & $0.4800^{+0.0198}_{-0.0191}$ & $0.4943^{+0.0279}_{-0.0282}$\\
$w$ & -- & $-1.0168^{+0.0281}_{-0.0284}$ & -- & -- & --\\
$\gamma$ & -- & -- & -- & -- & $0.0144^{+0.0100}_{-0.0090}$\\
$\xi_{DM}$ & -- & -- & -- & -- & $0.0143^{+0.0029}_{-0.0025}$\\
$\eta$ & -- & -- & -- & $-0.1361^{+0.0075}_{-0.0071}$ & --\\
\bottomrule
\end{tabular}
\end{adjustbox}
\end{table}
\begin{table}[!htbp]\centering\small
\caption{Joint numerical likelihood maxima for D1.}\label{tab:mleD1}
\begin{adjustbox}{max width=\linewidth}\begin{tabular}{lccccc}
\toprule
Parameter & $\Lambda$CDM & $w$CDM & HHDE & Model 1.7 & Model 5.6\\
\midrule
$M$ & $-19.3829$ & $-19.3900$ & $-19.4164$ & $-19.3916$ & $-19.3915$\\
$H_0$ & $68.958$ & $68.862$ & $68.315$ & $68.832$ & $68.830$\\
$100\omega_b$ & $1.3556$ & $9.3178$ & $4.9063$ & $0.5001$ & $5.2760$\\
$\Omega_{c0}$ & $0.27302$ & $0.12345$ & $0.28831$ & $0.31587$ & $0.46198$\\
$10^5\omega_r$ & $4.5946$ & $4.5946$ & $3.7592$ & $3.7610$ & $4.2334$\\
$\alpha$ & -- & -- & $0.00100$ & $0.36783$ & $0.11205$\\
$w$ & -- & $-1.06096$ & -- & -- & --\\
$\gamma$ & -- & -- & -- & -- & $0.19112$\\
$\xi_{DM}$ & -- & -- & -- & -- & $0.09047$\\
$\eta$ & -- & -- & -- & $-0.30000$ & --\\
\bottomrule
\end{tabular}
\end{adjustbox}
\end{table}
\begin{table}[!htbp]\centering\small
\caption{Joint numerical likelihood maxima for D4 within the stated domains. The dagger refers to the associated sample diagnostic in Table~\ref{tab:diagnostics}.}\label{tab:mleD4}
\begin{adjustbox}{max width=\linewidth}\begin{tabular}{lccccc}
\toprule
Parameter & $\Lambda$CDM & $w$CDM & HHDE & Model 1.7 & Model 5.6$^{\dagger}$\\
\midrule
$M$ & $-19.3802$ & $-19.3749$ & $-19.2783$ & $-19.3503$ & $-19.3471$\\
$H_0$ & $69.053$ & $69.363$ & $74.508$ & $71.132$ & $71.249$\\
$100\omega_b$ & $2.2534$ & $2.2509$ & $2.2306$ & $2.2367$ & $2.2366$\\
$\Omega_{c0}$ & $0.25376$ & $0.25184$ & $0.22272$ & $0.24005$ & $0.24018$\\
$10^5\omega_r$ & $4.2298$ & $4.2250$ & $4.0532$ & $4.2504$ & $3.7592$\\
$\alpha$ & -- & -- & $0.02059$ & $0.23725$ & $0.25074$\\
$w$ & -- & $-1.01607$ & -- & -- & --\\
$\gamma$ & -- & -- & -- & -- & $0.02747$\\
$\xi_{DM}$ & -- & -- & -- & -- & $0.01797$\\
$\eta$ & -- & -- & -- & $-0.13662$ & --\\
\bottomrule
\end{tabular}
\end{adjustbox}
\end{table}

\subsection{Probe combinations and an interaction comparison}\label{appen.additional}\label{appen.model17}

Table~\ref{tab:additionalconstraints} gives the HHDE constraints for all four data combinations.
\begin{table}[!htbp]\centering\small
\caption{HHDE marginal summaries under D1--D4. D2 and D4 use predicted sound horizons; D3 and D4 include the conditional Planck distance prior.}\label{tab:additionalconstraints}
\begin{adjustbox}{max width=\linewidth}\begin{tabular}{lcccc}
\toprule
Parameter & D1 & D2 & D3 & D4\\
\midrule
$M$ & $-19.416^{+0.053}_{-0.055}$ & $-19.326^{+0.050}_{-0.051}$ & $-19.411^{+0.052}_{-0.051}$ & $-19.273^{+0.048}_{-0.046}$\\
$H_0$ & $68.31^{+1.83}_{-1.83}$ & $72.31^{+1.69}_{-1.70}$ & $68.48^{+1.80}_{-1.69}$ & $74.70^{+1.65}_{-1.56}$\\
$100\omega_b$ & $4.395^{+3.279}_{-2.672}$ & $1.221^{+0.432}_{-0.444}$ & $2.236^{+0.015}_{-0.015}$ & $2.230^{+0.015}_{-0.015}$\\
$\Omega_{c0}$ & $0.2117^{+0.0808}_{-0.0670}$ & $0.2489^{+0.0227}_{-0.0296}$ & $0.3132^{+0.0158}_{-0.0156}$ & $0.2230^{+0.0036}_{-0.0035}$\\
$10^5\omega_r$ & $4.175^{+0.290}_{-0.287}$ & $4.121^{+0.306}_{-0.256}$ & $4.074^{+0.177}_{-0.163}$ & $4.074^{+0.177}_{-0.164}$\\
$\Omega_{M0}$ & $0.3226^{+0.0506}_{-0.0671}$ & $0.2733^{+0.0282}_{-0.0374}$ & $0.3609^{+0.0170}_{-0.0167}$ & $0.2630^{+0.0033}_{-0.0032}$\\
$\alpha$ & $0.1794^{+0.1719}_{-0.1255}$ & $0.1258^{+0.1183}_{-0.0889}$ & $0.0826^{+0.0091}_{-0.0095}$ & $0.0204^{+0.0057}_{-0.0057}$\\
$\beta$ & $0.4954^{+0.0777}_{-0.1046}$ & $0.6004^{+0.0610}_{-0.0812}$ & $0.5563^{+0.0250}_{-0.0247}$ & $0.7165^{+0.0064}_{-0.0065}$\\
\bottomrule
\end{tabular}
\end{adjustbox}
\end{table}

Model 1.4, with $\theta=\gamma$ and zero viscosity, provides a phenomenological comparison. Its local endpoint is $(1+\gamma,-\gamma,0)$, so negative $\gamma$ can support a locally physical future attractor. At the median parameter vectors from its distance fits, the coupling is negative and the background DM density becomes negative toward the early universe, violating the past-density condition. Table~\ref{tab:optional14} gives finite-sample summaries; the D4 acoustic quantities use the approximate drag-redshift prescription.
\begin{table}[!htbp]\centering\small
\caption{Phenomenological distance fits for Model 1.4, with the past-density violation and approximate acoustic calibration stated in the text. Here $n_w$ is the number of walkers, $n$ the number of steps per walker, $n_{\rm burn}$ the discarded initial steps, and $\tau$ the integrated autocorrelation time.}\label{tab:optional14}
\begin{tabular}{lcc}
\toprule
Parameter & D1 & D4\\
\midrule
$M$  &  $-19.391^{+0.057}_{-0.057}$  &  $-19.353^{+0.041}_{-0.045}$ \\
$H_0$  &  $68.85^{+1.89}_{-1.86}$  &  $71.24^{+1.41}_{-1.49}$ \\
$100\omega_b$  &  $5.138^{+3.252}_{-3.154}$  &  $2.237^{+0.015}_{-0.015}$ \\
$\Omega_{c0}$  &  $0.3081^{+0.1252}_{-0.1129}$  &  $0.2507^{+0.0043}_{-0.0043}$ \\
$10^5\omega_r$  &  $4.169^{+0.288}_{-0.277}$  &  $4.363^{+0.150}_{-0.190}$ \\
$\alpha$  &  $0.1686^{+0.1748}_{-0.1196}$  &  $0.1137^{+0.0208}_{-0.0211}$ \\
$\gamma$  &  $-0.0914^{+0.0660}_{-0.0731}$  &  $-0.0083^{+0.0023}_{-0.0022}$ \\
$n_w$  &  40  &  40 \\
$n$  &  20000  &  20000 \\
$n_{\rm burn}$  &  4000  &  4000 \\
$\widehat R_{\max}$  &  1.014  &  1.010 \\
$\mathrm{ESS}_{\min}$  &  3074  &  4182 \\
$[(n-n_{\rm burn})/\tau]_{\min}$  &  85.4  &  135.6 \\
\bottomrule
\end{tabular}

\end{table}
\FloatBarrier

\FloatBarrier
\bibliographystyle{JHEP}
\bibliography{refs}

\end{document}